\documentclass[a4paper,11pt]{article}
\pdfoutput=1 

\usepackage{jheppub} 
\usepackage{amsmath,amssymb,amsthm,amscd,graphicx}
\input epsf.sty

\theoremstyle{definition}

\newcommand{\CF}{{\cal F}}

\newcommand{\CJ}{{\cal J}}

\newcommand{\CN}{{\cal N}}
\newcommand{\CO}{{\cal O}}

\newcommand{\CQ}{{\cal Q}}

\newcommand{\CS}{{\cal S}}

\newcommand{\CV}{{\cal V}}

\def\IZ{{\mathbb Z}}
\def\IR{{\mathbb R}}
\def\IC{{\mathbb C}}

\newcommand{\re}{{\rm e}}
\newcommand{\ri}{{\rm i}}
\newcommand{\rd}{{\rm d}}

\newcommand{\mm}{\mathsf{p}}

\newcommand{\mH}{\mathsf{H}}

\newcommand{\mq}{\mathsf{q}}

\newcommand{\be}{\begin{equation}}
\newcommand{\ee}{\end{equation}}
\newcommand{\ba}{\begin{aligned}}
\newcommand{\ea}{\end{aligned}}
\newcommand{\ben}{\begin{eqnarray}\displaystyle}
\newcommand{\een}{\end{eqnarray}}

\newcommand{\sectiono}[1]{\section{#1}\setcounter{equation}{0}}

\newdimen\tableauside\tableauside=1.0ex
\newdimen\tableaurule\tableaurule=0.4pt
\newdimen\tableaustep
\def\phantomhrule#1{\hbox{\vbox to0pt{\hrule height\tableaurule width#1\vss}}}
\def\phantomvrule#1{\vbox{\hbox to0pt{\vrule width\tableaurule height#1\hss}}}
\def\sqr{\vbox{%
  \phantomhrule\tableaustep
  \hbox{\phantomvrule\tableaustep\kern\tableaustep\phantomvrule\tableaustep}%
  \hbox{\vbox{\phantomhrule\tableauside}\kern-\tableaurule}}}
\def\squares#1{\hbox{\count0=#1\noindent\loop\sqr
  \advance\count0 by-1 \ifnum\count0>0\repeat}}
\def\tableau#1{\vcenter{\offinterlineskip
  \tableaustep=\tableauside\advance\tableaustep by-\tableaurule
  \kern\normallineskip\hbox
    {\kern\normallineskip\vbox
      {\gettableau#1 0 }%
     \kern\normallineskip\kern\tableaurule}%
  \kern\normallineskip\kern\tableaurule}}
\def\gettableau#1{\ifnum#1=0\let\next=\null\else
\squares{#1}\let\next=\gettableau\fi\next}

\newcommand{\figref}[1]{Fig.~\protect\ref{#1}}

\title{\boldmath TBA equations and resurgent Quantum Mechanics}

\author[a]{Katsushi Ito,}
\affiliation[a]{Department of Physics, Tokyo Institute of Technology,\\ Tokyo, 152-8551, Japan\\}

\author[b]{Marcos~Mari\~no,}
\affiliation[b]{D\'epartement de Physique Th\'eorique \& Section de Math\'ematiques,\\ Universit\'e de Gen\`eve, Gen\`eve, CH-1211 Switzerland\\}

\author[a]{Hongfei Shu\,}

\abstract{We derive a system of TBA equations governing the exact WKB periods in one-dimensional 
Quantum Mechanics with arbitrary polynomial potentials. These equations provide a generalization of the ODE/IM correspondence, and they can be 
regarded as the solution of a Riemann--Hilbert problem in resurgent Quantum Mechanics formulated by Voros. Our derivation builds upon the solution of similar 
Riemann--Hilbert problems in the study of BPS spectra in $\CN=2$ gauge theories and of minimal surfaces in AdS. We also show that our TBA equations, combined 
with exact quantization conditions, provide a powerful method to solve spectral problems in Quantum Mechanics. We illustrate our general analysis with a detailed study 
of PT-symmetric cubic oscillators and quartic oscillators. }    

\begin{document}
\maketitle
\flushbottom
\sectiono{Introduction}

In quantum theories one typically obtains a plethora of perturbative series, associated to different 
classical configurations. These series are the basic building block in many 
computations, but they are almost always divergent. The theory of resurgence provides a way to organize these 
series into a coherent picture of the quantum system. First 
of all, the divergent series are converted into meaningful objects by doing Borel resummations in the perturbation parameter. 
Second, one discovers that the different series are in fact intimately related. One way 
to understand this relation is through the Stokes phenomenon, which simply states that Borel resummations are discontinuous along 
specific rays in the complex plane. It turns out that, given one specific series, the discontinuities of its Borel transform encode the information about other series in the problem. 

The resurgent program was applied to Quantum Mechanics in one dimension through an ``exact" 
version of the WKB method \cite{bpv,voros,voros-quartic,dpham, reshyper, ddpham}. In this 
context, the different perturbative series are the WKB series around different classical trajectories, which we will call WKB or quantum 
periods. This led to many powerful results, 
like for example exact connection formula and exact quantization conditions for spectral problems. In the case of the pure quartic oscillator, 
Voros found in \cite{voros-quartic} a complete determination of the structure of the discontinuities 
for the different WKB periods involved in the problem. However, this flourishing of ideas did not provide alternative computational methods in Quantum Mechanics. 
One was supposed to calculate the WKB series and their Borel resummations with traditional tools. 

Partially inspired by the results of \cite{voros-quartic}, a correspondence between certain quantum mechanical models in one dimension 
and integrable models (or ODE/IM correspondence) 
was proposed by Dorey and Tateo in \cite{dt}. This correspondence was originally based on functional relations discovered in the context of 
resurgent Quantum Mechanics, which are similar to Baxter-type equations appearing in integrable systems. The ODE/IM correspondence 
makes it possible to write down TBA equations which calculate efficiently and exactly some of the quantities appearing in the quantum mechanical models, like 
spectral determinants and Borel resummations of WKB periods. An important limitation of the ODE/IM correspondence is that it applies to very special quantum-mechanical 
models, namely, monic potentials of the form $V(x)= x^M$ (and a limited amount of perturbations thereof).

In this paper we will present TBA equations governing the WKB periods for general polynomial potentials in one dimension, providing in this way 
a generalization of the ODE/IM correspondence. The basic idea was already pointed out by Voros in his seminal 
paper \cite{voros-quartic}, where he called it the ``analytic bootstrap" for the exact WKB method. In this approach, 
the fundamental objects are the WKB periods, which are (Borel resummed) perturbative series in the Planck 
constant. These periods can be characterized by two types of data: their {\it classical limit} and their {\it discontinuity structure}, 
which has been known since the work 
of \cite{voros-quartic,dpham, ddpham}. In the theory of resurgence, this discontinuity structure 
is encoded in the action of the so-called Stokes automorphisms. 
As Voros explained, we can think about these data as defining a Riemann--Hilbert problem. 
Building on recent developments in seemingly very different contexts 
\cite{ks,gmn,gmn2,alday-mal,aldaygm,y-system,hito, tateo-extended}, we show that Voros' Riemann--Hilbert problem has a solution in terms of a 
TBA-like system, which determines the exact dependence of the WKB periods 
on the Planck constant\footnote{A different solution to the problem was proposed in \cite{voros-fp} in terms of 
a fixed-point equation. It is likely that this solution is closely related to the TBA system proposed here.}. 
The TBA system of \cite{dt} is recovered as a particular case of our general story, as we fine tune the moduli of the potential to a point with enhanced symmetry. 
From the point of view of the theory of resurgence, the TBA system we obtain here provides an efficient way to calculate and 
resum the perturbative expansion of the WKB periods in a single strike. 

The connection between Voros' Riemann--Hilbert problem for the exact WKB method, and the Riemann--Hilbert problems appearing in 
\cite{cv,gmn,gmn2, alday-mal,aldaygm, y-system,hito}, has been already pointed out in the literature. In particular, Gaiotto argued in \cite{oper} that the 
conformal limit of the TBA equations for Hitchin systems obtained in \cite{gmn,gmn2} leads to TBA equations solving 
Schr\"odinger problems, and he presented some examples going beyond the conventional ODE/IM 
correspondence (see \cite{cdz,dumitrescu,ito-shu,bridgeland} for related discussions). In this paper we put the results of \cite{oper} in the 
context of the general theory of 
resurgence, and we derive the TBA equations governing general polynomial potentials in one dimension. In addition, 
we provide detailed tests and illustrations of the resulting formalism: we use the TBA equations to compute exact WKB periods, and we extract the perturbative 
expansion from the large $\theta$ expansion of the TBA system. We also use the TBA system to compute the spectrum of various interesting quantum-mechanical 
systems, like PT-symmetric cubic oscillators and quartic oscillators. 

We would like to emphasize that the ``resurgent" perspective on the construction of the TBA system 
might have many interesting applications in more general contexts. From the point of view of 
the theory of resurgence, the data for the Riemann--Hilbert problem (the classical limit of the quantum objects and the 
structure of their Stokes automorphisms) are quite universal, and characterize 
many different quantum systems. 
This opens the possibility to reformulate and solve quantum theories uniquely in terms of these data, instead of using path integrals 
or Schr\"odinger-type equations\footnote{This point of view has been advocated in particular by M. Kontsevich.}. 

This paper is organized as follows. In section \ref{sec-wkb} we review some results on the exact WKB 
method and the theory of resurgence which will be useful in the paper. 
In section \ref{sec-tba} we derive the TBA system. We present two different derivations: the first 
one gives the TBA system as a solution to a Riemann--Hilbert problem, while the 
second one is based on an analysis of the Schr\"odinger equation along the lines of \cite{dt-stokes,y-system}. 
We also present some general properties of the resulting TBA 
system and we discuss the wall-crossing structure. In sections \ref{sec-cubic} and \ref{sec-quartic} we 
apply our general formalism to two important examples in Quantum Mechanics, namely the cubic and the quartic 
oscillators. We show in particular how to use our formalism to solve concrete spectral problems. Finally, in section \ref{sec-conclusions} 
we state our conclusions and list some interesting open problems. 
The Appendix \ref{app-ha} contains some details on the perturbative calculation of quantum periods. 

\sectiono{The exact WKB method and resurgent Quantum Mechanics}
\label{sec-wkb}
In this section we give a lightning review of the exact WKB method from the point of view of the theory of resurgence. This method, in its resurgent incarnation, 
was developed in e.g. \cite{voros,voros-quartic,reshyper,dpham, ddpham,unfolding}, 
and it has been recently reviewed in \cite{in-exactwkb}. For more general reviews of the theory of 
resurgence, see \cite{mmlargen,abs}. 

Let us consider the stationary Schr\"odinger equation for a non-relativistic particle in a potential $V(q)$ and with energy $E$:
\be
\label{schrodinger}
-\hbar^2 \psi''(q) + (V(q)-E) \psi(q)=0. 
\ee
The standard WKB method produces asymptotic expansions in $\hbar$ for the solutions to this equation. This goes as follows. 
Let us consider the following ansatz
\be
\label{yansatz}
\psi(q)=\exp \left[ {\ri \over \hbar} \int^q Q(q') \rd q' \right]. 
\ee
Then, when plugged in (\ref{schrodinger}), we obtain a Riccati equation for $Q(q)$, 
\be
\label{riccati}
 Q^2(q)-\ri \hbar\frac{\rd Q(q)}{\rd q}=p^2(q), \quad p(q)=(E-V(q))^{1/2},
\end{equation} 
which can be solved in power series in $\hbar$:
\be
\label{yps}
Q(q)=\sum_{k=0}^\infty Q_k(q)\hbar^k. 
\ee
The functions $Q_k$ can be calculated recursively. If we split the formal power series in (\ref{yps}) into even and odd powers of $\hbar$, 
 \be
 \label{yyp}
 Q(q) =  P(q)+Q_{\rm odd} (q),
\ee
one finds that $Q_{\rm odd}(q)$ is a total derivative, 
\be
Q_{\rm odd}(q)={\ri \hbar \over 2} {\rd \over \rd q} \log  P(q).
\ee
The wavefunction (\ref{yansatz}) can then be written as 
\be
\psi(q) ={1\over {\sqrt{ P(q)}}} \exp \left[ {\ri \over \hbar} \int^q P(q') \rd q' \right]. 
\ee
The formal power series $P(q)$ has the structure,  
\be
P(q) = \sum_{n \ge 0} p_n(q) \hbar^{2n}, 
\ee
where $p_0(q) =p(q)$ is the classical momentum. Geometrically, we can regard $P(q) \rd q$ as a meromorphic differential on the curve defined by 
\be
\label{wkb-curve} 
y^2 = 2(E- V(q)). 
\ee
Note that if $V(q)$ is a polynomial potential, as we will assume in this paper, then (\ref{wkb-curve}) is a hyperelliptic curve defining a 
Riemann surface $\Sigma_{\rm WKB}$, which we will call {\it WKB curve}. This curve depends on a set of moduli which include the 
energy $E$ and the parameters of the potential $V(q)$. 

The basic objects in the exact WKB method are the periods of $P(q) \rd q$ along the one-cycles $\gamma \in H_1(\Sigma_{\rm WKB})$, 
which we will call {\it WKB periods} or {\it quantum periods}. We will denote them as 
\be
\Pi_\gamma(\hbar)= \oint_\gamma P(q) \rd q, \qquad \gamma \in H_1(\Sigma_{\rm WKB}).
\ee
A related quantity is the so-called {\it Voros multiplier} or {\it Voros symbol}, which is simply the exponent of the WKB period:
\be
\CV_\gamma= \exp\left( {\ri \over \hbar} \Pi_\gamma\right). 
\ee
As we will see, the algebra of Stokes discontinuities has a simpler expression in terms of Voros multipliers.    
The WKB periods are formal power series in even powers of $\hbar$, as $P(q)$, 
\be
\label{pi-wkb-expansion}
\Pi_\gamma(\hbar)=\sum_{n \ge 0} \Pi^{(n)}_\gamma \hbar^{2n}, \qquad  \Pi^{(n)}_\gamma = \oint_\gamma p_n(q) \rd q. 
\ee
Note that the coefficients $ \Pi^{(n)}_\gamma$ depend on the moduli of the WKB curve. The calculation of these coefficients 
at high order can be quite involved, even for simple quantum systems. In Appendix \ref{app-ha} we summarize two powerful 
methods to calculate these quantum corrections. 

In the exact WKB method, the quantum periods, which start their life as formal power series in $\hbar$, are promoted to actual functions 
by the procedure of Borel resummation. It turns out that, for fixed values of the moduli of the curve, the series (\ref{pi-wkb-expansion}) has a doubly-factorial 
growth, 
\be
\Pi^{(n)}_\gamma \sim (2n)!. 
\ee
The Borel transform of the quantum period is then defined as 
\be
\widehat \Pi_\gamma(\xi) = \sum_{n \ge 0} {1\over (2n)!} \Pi^{(n)}_\gamma \xi^{2n}, 
\ee
and it is analytic in a neighbourhood of the origin in the $\xi$-plane. Moreover, it can be analytically continued to a function 
on the complex plane, displaying in general various types of singularities (typically poles and branch cuts.)
The Borel resummation of the quantum period is defined by a Laplace transform, 
\be
\label{borel-transform}
s\left( \Pi_\gamma\right)(\hbar)= {1\over \hbar} \int_0^\infty \re^{-\xi/\hbar} \widehat \Pi_\gamma(\xi) \rd \xi, \qquad \hbar \in \IR_{>0}. 
\ee
If this integral converges  when $\hbar$ is small enough, the quantum period is said to be {\it Borel summable}. 
We will however need a more general definition of Borel resummation along a direction in the complex plane, specified by an angle $\varphi$:
\be
\label{bt-theta}
s_\varphi \left( \Pi_\gamma\right)(\hbar)= {1\over \hbar} \int_0^{\re^{\ri \varphi} \infty} \re^{-\xi/\hbar} \widehat \Pi_\gamma(\xi) \rd \xi. 
\ee
Let us note that $\Pi_\gamma$ is Borel summable along the direction $\varphi$ if and only if the series 
$\Pi_\gamma(\re^{\ri \varphi} \hbar)$ is Borel summable. 
Therefore, considering general directions for the Borel resummation is equivalent to considering
different rays in the complex $\hbar$-plane. 

One of the basic facts of the theory of Borel resummation is that {\it singularities} of the Borel transform lead to 
obstructions to Borel summability. For example, if $\widehat \Pi_\gamma$ 
has a singularity on the positive real axis, then (\ref{borel-transform}) is not defined. More precisely, 
singularities lead to {\it discontinuities} in the Borel resummation: if $\widehat \Pi_\gamma$ has 
a singularity along a direction $\varphi$ in the Borel plane of the variable $\xi$, there will be a jump in the 
directional Borel resummation (\ref{bt-theta}) as we cross the angle $\varphi$. To measure this discontinuity, it is 
useful to introduce the {\it lateral Borel resummations} along a direction $\varphi$: 
\be
s_{\varphi \pm} (\Pi_\gamma)\left(\re^{\ri \varphi}\hbar\right)= \lim_{\delta \rightarrow 0} \,  
s \left( \Pi_\gamma\right)\left(\re^{\ri \varphi\pm \ri \delta}\hbar \right), \qquad \hbar \in \IR_{>0}, 
\ee
which are the Borel resummations just after and just before crossing the discontinuity. 
Lateral Borel resummations can be equivalently defined in terms of Borel resummations (\ref{bt-theta}) 
along the directions $\varphi \pm \delta$. The discontinuity across the direction $\varphi$ is now defined by, 
\be
{\rm disc}_\varphi\,  \left( \Pi_\gamma\right)= s_{\varphi +} (\Pi_\gamma)-s_{\varphi -} (\Pi_\gamma). 
\ee
We can think of $s_{\varphi \pm}$ as operators, acting on formal power series. One then defines the {\it Stokes automorphism} in the direction $\varphi$ as 
\be
\mathfrak{S}_\varphi = s_{\varphi+} \circ s_{\varphi-}^{-1}.  
\ee
In this definition we follow the convention in \cite{ddpham} 
(the opposite convention, which amounts to exchange $\mathfrak{S}_\varphi$ with $\mathfrak{S}_\varphi^{-1}$, is used in 
e.g. \cite{in-exactwkb}). 
One of the key points in the resurgent structure of Quantum Mechanics is that the operators $s_{\varphi \pm}$ and 
$\mathfrak{S}_\varphi$, when acting on a quantum period, are given by functions involving 
just the other quantum periods. In other words, the structure of discontinuities in Quantum Mechanics involves just the 
quantum periods on the WKB curve, and no other formal power series intervene. 

\begin{figure}[tb]
\begin{center}
\resizebox{60mm}{!}{\includegraphics{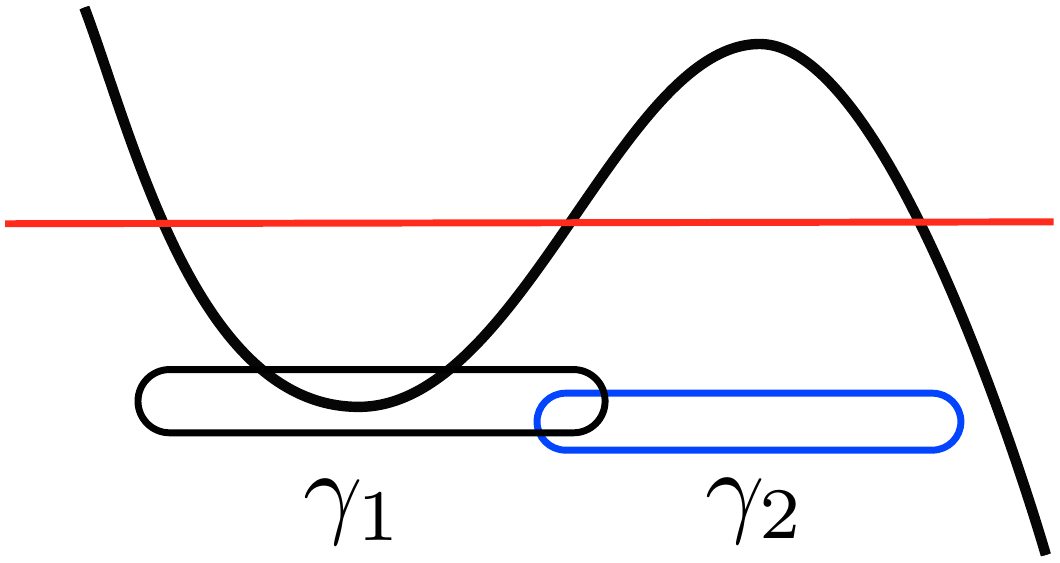}}
\end{center}
  \caption{The WKB cycles for the Delabaere--Pham formula.}
\label{dpham-fig}
\end{figure}

Let us give an important example of such a structure: the {\it Delabaere--Pham formula} (Theorem 2.5.1 of \cite{dpham}, see also 
Appendix A of \cite{in-exactwkb}). Let us 
consider the quantum periods associated to the two WKB cycles 
shown in \figref{dpham-fig}: one of the cycles, $\gamma_1$, corresponds to a classically allowed interval, 
while the cycle $\gamma_2$ corresponds to a 
classically forbidden interval (in the language of Stokes graphs \cite{in-exactwkb}, $\gamma_2$ encircles a regular saddle trajectory). 
The Delabaere--Pham formula states that the quantum period $\Pi_{\gamma_1}$ is 
{\it not} Borel summable along the direction $\varphi=0$, and therefore there will 
be a discontinuity as we cross it. The discontinuity can be expressed in terms of the quantum period $\Pi_{\gamma_2}$ (which is Borel 
summable), and it is given by
\be
\label{dpf-1}
{\rm disc} \left( \Pi_{\gamma_1}\right)(\hbar)= -\ri \hbar  \log\left( 1+ \exp\left(- {\ri   \over \hbar}\Pi_{\gamma_2} (\hbar)\right) \right). 
\ee
(When no angle is indicated in a discontinuity formula, it is understood that it refers to $\varphi=0$). In calculating the quantum period 
associated to the cycle $\gamma_2$ around the forbidden interval, we have 
to make a choice of branch cut of the momentum function $p(q)$. The choice is such that $\ri \Pi_{\gamma_2}$ is real and positive, 
therefore the discontinuity appearing in the r.h.s. of (\ref{dpf-1}) is 
exponentially small. In other words, it is a non-perturbative effect in $\hbar$. 
The Delabaere--Pham formula can be elegantly expressed in terms of Stokes automorphism and Voros multipliers, and it reads:
\be
\label{dpham-formula}
\mathfrak{S} \, \CV_{\gamma_1}=\left (1+ \CV^{-1}_{\gamma_2} \right)\CV_{\gamma_1}.
\ee
Here we have assumed that there is a single cycle $\gamma_2$ associated to a forbidden interval and 
intersecting the cycle $\gamma_1$. When there are more cycles $\gamma_a$, $a=2, \cdots, r$, associated to 
forbidden intervals and intersecting $\gamma_1$, the above formula generalizes to 
\be
\mathfrak{S} \, \CV_{\gamma_1} =\prod_{a=2}^{r} \left (1+ \CV_{\gamma_a}^{-1} \right)^{\langle \gamma_a, \gamma_1 \rangle} \CV_{\gamma_1}, 
\ee
where $\langle \gamma_a, \gamma_1 \rangle$ is the intersection of the cycles, which can be appropriately 
defined after careful choices of orientation 
(see \cite{dpham} for the details). 

The Delabaere--Pham formula (\ref{dpham-formula}) provides a simple example of a resurgent 
structure in Quantum Mechanics: the discontinuity of the 
Borel resummation of the quantum period $\Pi_{\gamma_1}$ is completely encoded in another quantum period $\Pi_{\gamma_2}$. 
Note that the discontinuity of $\Pi_{\gamma_1}$ is a non-perturbative effect, but at the same time 
it is determined by its all-orders perturbative structure through the Borel transform. 

One of the goals of the resurgent analysis of Quantum Mechanics is a complete determination of the 
discontinuity structure of the quantum periods (or equivalently, of the Voros multipliers). 
Such a result was obtained by Voros in \cite{voros-quartic} in the case of the pure quartic oscillator in 
one dimension, and further results in one-dimensional 
Quantum Mechanics were obtained in \cite{dpham,ddpham} and more recently in \cite{in-exactwkb}. 
As we will see, this discontinuity structure will be a crucial ingredient in our approach. 

The other goal of resurgent Quantum Mechanics is to find exact quantization 
conditions (EQCs) for spectral problems. The EQC is typically obtained as the condition that a WKB solution decreasing along one ray in the complex plane 
remains decreasing when it is continued along a different ray. To obtain these conditions, we have to solve the ``connection problem" of relating 
WKB solutions in different regions. The correct solution of the connection problem was found by Voros in \cite{voros, 
voros-quartic}, and independently by Silverstone in \cite{silverstone}. Using the Voros--Silverstone 
connection formula, one can systematically derive EQCs for many problems in one-dimensional 
Quantum Mechanics. As emphasized in \cite{bpv}, they involve in principle {\it all} the (Borel resummed) Voros multipliers  
associated to the WKB curve, and they typically have the form of a single functional relation between them: 
\be
\label{eqc-abstract}
\CQ\left( \CV_{\gamma_1}, \cdots, \CV_{\gamma_{r}} \right)=0. 
\ee
When the Voros multipliers are not Borel summable in the standard sense, one has to use lateral 
Borel resummations, and the form of the EQC depends on the 
choice of lateral resummation. Different forms of the EQC are then related by the corresponding 
Stokes automorphisms. In this paper we will analyze EQCs in some concrete 
examples and we will rely on the results obtained in \cite{voros, voros-quartic, dpham, 
ddpham,unfolding}\footnote{Historically, some of these EQCs were obtained in the framework of 
instanton calculus, see e.g. \cite{zinn-justin, zjj1, zjj2}. An alternative approach to 
derive EQCs is the uniform WKB method, see e.g. \cite{alvarez,alvarez-casares2, alvarez-casares,dunne-unsal}}.

\begin{figure}[tb]
\begin{center}
\resizebox{75mm}{!}{\includegraphics{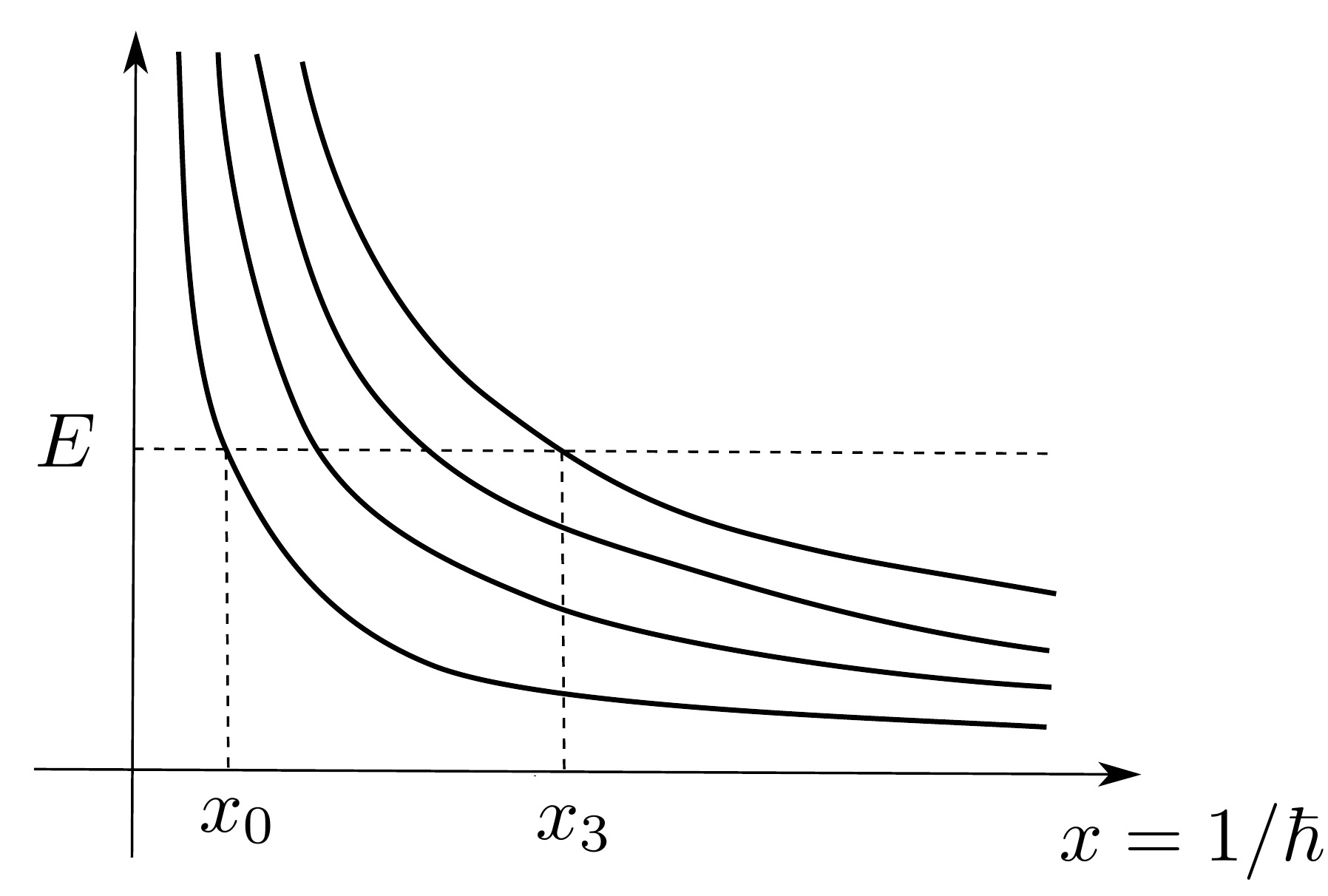}}
\end{center}
  \caption{The spectrum of a quantum system can be regarded as a discrete series of 
  energies for a fixed value of $\hbar$, or conversely, as a discrete 
  series of values of the inverse Planck constants $x=\hbar^{-1}$ for a fixed value of the energy $E$. 
  In this figure we show the very first energy levels $E_n$, as functions of $x$, and 
  the sequence of $x_n$ corresponding to the energy $E$. 
}
\label{energyhbar}
\end{figure}

Let us make a final remark on the use of EQCs like (\ref{eqc-abstract}). The Voros 
multipliers depend on $\hbar$ and on the moduli of the WKB curve $\Sigma_{\rm WKB}$, 
in particular on the energy $E$. Assuming that the spectral problem leads to a 
discrete set of energies (or resonances), the
EQCs define a discrete, infinite family of codimension one submanifolds in 
the ``extended" moduli space parametrized by $\hbar$, $E$ and the
additional parameters of the potential. 
In particular, if we fix these additional parameters, we obtain a discrete, infinite family of 
curves in the $(E, \hbar)$ plane, 
labelled by the quantum number $n=0, 1, 2, \cdots$. Usually, one thinks of these 
curves as the discrete values of the energy of the system $E_n(\hbar)$, for a fixed value of $\hbar$. 
However, one can reverse this relation and think of the spectrum of a Hamiltonian as a series of discrete values $x_n(E)$ of the {\it inverse Planck constant}
\be
x={1\over \hbar},
\ee
for a fixed value of the energy $E$. By construction, the two spectra $E_n(\hbar)$ and $x_n(E)$ are related by
\be
\label{Exrelation}
E_n \left({1\over x_n(E)} \right)= E.
\ee
This is illustrated in \figref{energyhbar}. This alternative view of the spectrum of a Hamiltonian 
in terms of the values $x_n(E)$ has been advocated by Voros (see e.g. \cite{voros-quartic}) and we will call 
it the {\it Voros spectrum} of the Hamiltonian. This is in fact the natural form of the spectrum when we think about the WKB periods 
as solutions of a Riemann--Hilbert problem. As we will see in this paper, our TBA equations, combined with EQCs, 
will give the Voros spectrum, rather than the traditional spectrum. 
We will derive the TBA equations satisfied by the quantum periods in the next section.

\sectiono{TBA equations and a generalized ODE/IM correspondence}
\label{sec-tba}

\begin{figure}[tb]
\begin{center}
\resizebox{75mm}{!}{\includegraphics{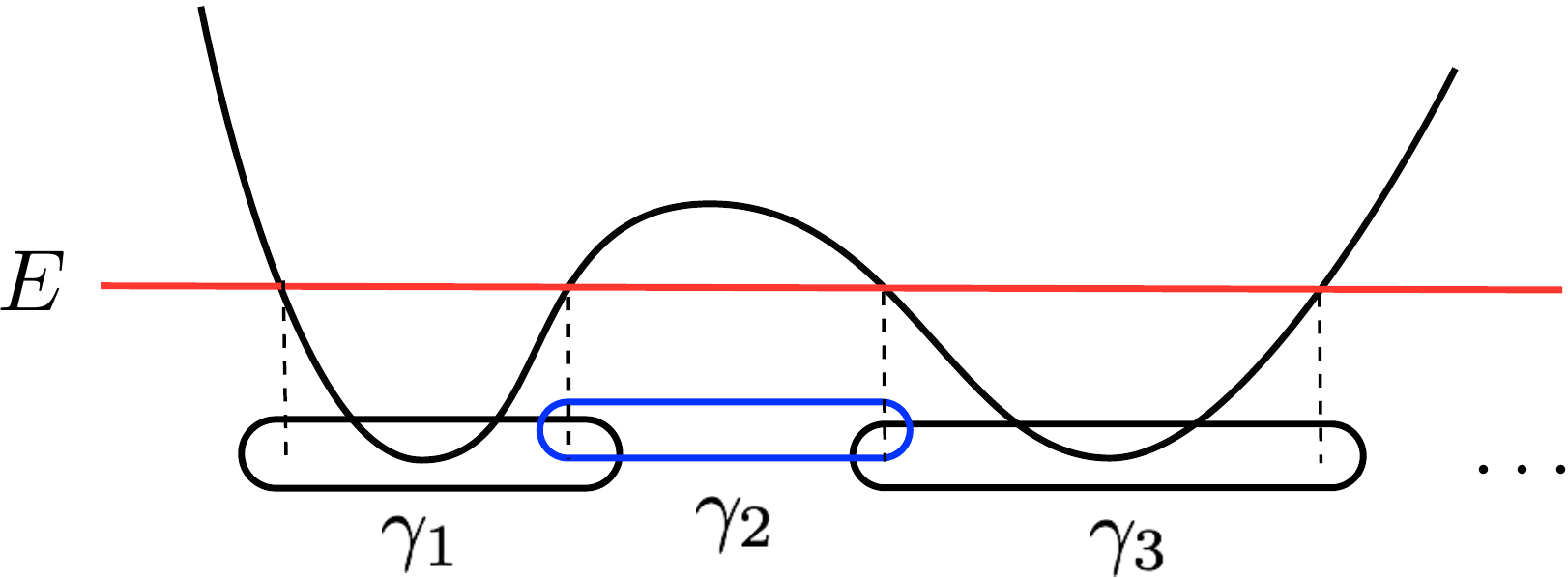}}
\end{center}
  \caption{To derive the TBA system obeyed by the WKB periods, we consider a region in moduli 
  space where the discontinuities of their Borel resummations are encoded by the Delabarere--Pham formula. 
  In this region all the turning points are real and different, and we have a sequence of cycles encircling classically 
  allowed and classically forbidden intervals in an alternate way.}
\label{minimal-fig}
\end{figure}

In this section we will derive the TBA equations obeyed by the quantum periods. We will provide two 
different derivations. In the first one, we will consider the Riemann--Hilbert problem for the periods stated by Voros in \cite{voros-quartic}, namely: 
given their classical limit $\Pi_\gamma^{(0)}$, and the discontinuities of their Borel-resummed versions, can we reconstruct the WKB periods exactly? Then, following 
\cite{gmn}, we will write a solution of this problem, in terms of this TBA system, which satisfies this data. The second derivation 
will be more constructive, in the spirit of \cite{y-system,ito-shu}. 

\subsection{TBA equations and Riemann-Hilbert problem}
\label{sec-derivation}

Let us consider a polynomial potential of degree $r+1$, $V_{r+1}(q)$. The WKB curve is a 
hyperelliptic curve describing a Riemann surface of genus $g=[r/2]$. Let us consider a region in the moduli space of the curve 
where all the turning points $q_i$, $i=1, \cdots, r+1$, are real and different. Without loss of generality, let us order the turning points as 
\be
q_1 <q_2 < \cdots <q_{r+1}. 
\ee
In addition, let us choose the sign of the term $q^{r+1}$ in the potential in such a way that the interval 
\be
[q_i, q_{i+1}], \qquad  i=1, \cdots, r, 
\ee
corresponds to a classically allowed interval when $i$ is odd, and to a classically 
forbidden interval when $i$ is even. We then have a situation like the one depicted in \figref{minimal-fig}. 
The cycle $\gamma_i$, $i=1, \cdots, r$, goes around the interval $[q_i, q_{i+1}]$. 
We choose the branch cuts and orientations of the cycles in such a way that 
\be
\label{masses}
m_{2i-1}= \Pi^{(0)}_{\gamma_{2i-1}}= 2 \int_{q_{2i-1}}^{q_{2 i}} p(q) \rd q, \qquad m_{2i}= \ri \, \Pi^{(0)}_{\gamma_{2i}}=2 \ri \int_{q_{2i}}^{q_{2i+1}}p(q) \rd q,  
\ee
are real and positive. Borrowing the language of two-dimensional integrable theories, we will often refer to these quantities as ``masses." 
The Delabaere--Pham formula says that the quantum periods $\Pi_{\gamma_{2i}}$ are Borel summable along the 
positive real axis of $\hbar$, while $\Pi_{\gamma_{2i-1}}$ are not. In addition, we have the following discontinuity formula, 
\be
\label{disc1}
{\rm disc}\, \Pi_{\gamma_{2i-1}}= -\ri \hbar  \log\left( 1+ \exp\left( -{\ri   \over \hbar}\Pi_{\gamma_{2i-2}} (\hbar)\right) \right) -\ri \hbar 
\log\left( 1+ \exp\left( -{\ri   \over \hbar}\Pi_{\gamma_{2i}} (\hbar)\right) \right), \qquad  
\ee
Equivalently, in terms of Stokes automorphisms and Voros multipliers, we have 
\be
\label{disc2}
\mathfrak{S} \, \CV_{\gamma_{2i-1}} = \left (1+ \CV^{-1}_{\gamma_{2i-2}} \right) \left (1+ \CV^{-1}_{\gamma_{2i}} \right)  \CV_{\gamma_{2 i-1}}. 
\ee
A similar discontinuity formula holds along the negative real axis (this already follows from the fact that the quantum periods are 
formal power series in {\it even} powers of $\hbar$.) However, 
there is an additional discontinuity in the complex plane of $\hbar$. The reason is that, as we 
rotate $\hbar \rightarrow \pm \ri \hbar$, $\hbar \in \IR$, classically allowed and classically 
forbidden intervals are exchanged. Along this direction we have a discontinuity formula like (\ref{disc1}), (\ref{disc2}), but now the 
odd cycles are exchanged with the even cycles. 
In order to provide a unified description of both types of discontinuities, we introduce the functions
\be
\label{def-eps}
\ba
-\ri \epsilon_{2i-1}\left( \theta+ {\ri \pi \over 2} \pm \ri \delta \right)&={1\over \hbar} s_{\pm} \left( \Pi_{\gamma_{2i-1}} \right)(\hbar),\\
-\ri \epsilon_{2i}\left( \theta\right)&={1\over \hbar} s\left( \Pi_{\gamma_{2i}} \right)(\hbar), 
\ea
\ee
where $0<\delta\ll 1$, and we identify 
\be
\re^{\theta}= {1\over \hbar}. 
\ee
The Delabaere--Pham discontinuities can now be encoded in a single formula,
\be
\label{dp-dg}
{\rm disc}_{\pi/2} \epsilon_a(\theta)= L_{a-1}(\theta) + L_{a+1}(\theta), \qquad a=1, \cdots, r, 
\ee
where
\be
\label{l-def}
L_a(\theta)= \log \left( 1+ \re^{-\epsilon_a(\theta)} \right), 
\ee
and with the understanding that $L_0=L_{r+1}=0$. A similar formulae holds for the discontinuity at $-\pi/2$. We now have to 
solve the Riemann--Hilbert problem for the functions $\epsilon_a (\theta)$, $a=1, \cdots,r$, 
which satisfy the discontinuity formula (\ref{dp-dg}) and have the asymptotic behavior
\be
\epsilon_a(\theta) - m_a \re^{\theta}=\CO(\re^{-\theta}), \qquad \theta \rightarrow \infty. 
\ee
The solution to this problem is given by the following TBA system 
\be
\label{gentba}
\epsilon_a(\theta)= m_a \re^\theta - \int_\IR {L_{a-1}(\theta') 
\over\cosh(\theta-\theta')} {\rd \theta' \over 2 \pi}- \int_\IR {L_{a+1}(\theta')
\over\cosh(\theta-\theta')} {\rd \theta' \over 2 \pi}, \qquad a=1, \cdots, r.  
\ee
When we expand (\ref{gentba}) at large $\theta$, we find the all-orders asymptotic expansion
\be
\label{wkb-expansion}
\epsilon_a(\theta)\sim  m_a \re^\theta+ \sum_{n \ge 1} m_a^{(n)} \re^{(1-2n) \theta}, 
\ee
where
\be
\label{qc-m}
m_a^{(n)}= {(-1)^n \over \pi} \int_\IR \re^{(2n-1) \theta} \left( L_{a-1}(\theta)+ L_{a+1}(\theta) \right) \rd \theta. 
\ee
In view of (\ref{def-eps}), the WKB expansion of the periods can be recovered from the above expansion as
\be
\label{mcorr-wkb}
m_{2i-1}^{(n)}= (-1)^n \Pi^{(n)}_{\gamma_{2i-1}}, \qquad m_{2i}^{(n)}=\ri \Pi^{(n)}_{\gamma_{2i}}. 
\ee
The TBA system (\ref{gentba}) is the ``conformal limit" of the TBA system considered in \cite{y-system,hito}, in which one replaces $m_a \cosh(\theta)$ by 
$m_a \re^{\theta}$. This could have been anticipated based on the results of \cite{oper}. In the above derivation we have chosen a particular region in moduli space. 
From the point of view of resurgence, this region is special in the sense that the discontinuities of the periods are captured by the Delabaere--Pham formula in a simple way. In \cite{toledo, toledo-thesis} this region 
is called the ``minimal chamber" because it leads to a minimal number of $\epsilon$ functions. 

\subsection{Generalized ODE/IM correspondence}
Let us now offer a more direct derivation of the TBA system, following the techniques in \cite{y-system}. 
We start from the Schr\"odinger equation for a potential of degree $r+1$, which 
we write as
\be
\label{eq:Schrodinger-rescaled}
\left(-\partial_{z}^{2}+z^{r+1}+\sum_{a=1}^{r}b_{a}z^{r-a}\right){\psi(z, b_a)}=0. 
\ee
Here $z$ is a complex variable, and $b_a$, $a=1,\cdots,r$, are complex coefficients. The case in which $b_1=\cdots=b_{r-1}=0$ and $b_r\neq 0$ is the one considered in the usual ODE/IM correspondence, where $b_r$ plays the r\^ole of spectral parameter. The derivation that follows can be regarded as a 
generalization of the techniques of \cite{blz-spectral, dt-stokes} to general coefficients $b_a$.

The equation (\ref{eq:Schrodinger-rescaled}) is invariant under the Symanzik rotation 
\be
\label{sym-rotation}
(z, b_a)\to (\omega z, \omega^{a+1}b_a), \qquad \omega=\re^{\frac{2\pi \ri}{r+3}}. 
\ee
Let us define the coefficients $B_m$ as
\be
 \left(1+\sum_{a=1}^rb_a z^{-a-1}\right)^{1/2}:= 1+\sum_{m=1}^{\infty}B_{m}z^{-m}.
 \ee
By using the WKB approximation, one finds that the fast decaying solution of (\ref{eq:Schrodinger-rescaled}) at infinity around the positive real axis behaves as \cite{Sibuya}
\be
y(z, b_a)\sim\frac{1}{\sqrt{2\ri}}z^{n_{r}}\exp\Big(-\frac{2}{r+3}z^{\frac{r+3}{2}}\Big),\label{eq:Decay-solution-S0}
\ee
where $n_r$ is defined by
\be
n_{r}=\begin{cases}
-\frac{r+1}{4} & \quad r+1:\mbox{odd}\\
-\frac{r+1}{4}-B_{\frac{r+3}{2}} & \quad r+1:\mbox{even}
\end{cases}
\ee
Under the Symanzik rotation we have
\begin{align}
B_{\frac{r+3}{2}}(\omega^{-k(a+1)}b_{a})=(-1)^{k}B_{\frac{r+3}{2}}(b_a).
\end{align}
Let us now consider the sector in the complex $z$-plane, 
\be
{\cal S}_k=\left\{ z \in \IC  :\left|\mbox{arg}(z)-\frac{2k\pi}{r+3}\right|<\frac{\pi}{r+3} \right\}.
\ee
The solution (\ref{eq:Decay-solution-S0}) is the subdominant solution at infinity in the sector $\CS_0$. By using the Symanzik rotation, we can 
obtain the subdominant solution in the sector $\CS_k$, as
\be
y_{k}(z, b_{a})=\omega^{\frac{k}{2}}y(\omega^{-k}z,\omega^{-(a+1)k}b_{a}).
\ee
As is well-known, the Wronskian of two solutions
\begin{align}
W_{k_1,k_2}(b_a)\equiv W[y_{k_1}(z,b_a), y_{k_2}(z,b_a)]=y_{k_1}(z,b_a)\partial_zy_{k_2}(z,b_a)-y_{k_2}(z,b_a)\partial_zy_{k_1},
\end{align}
is independent of $z$. We introduce the following notation for the shift of functions:
\begin{align}
f^{[j]}(z,b_{a})&:=f(\omega^{-j/2}z,\omega^{-j(a+1)/2}b_{a}).
\end{align}
It is easy to find that 
\begin{align}
W_{k_1+1, k_2+1}(b_a)=W_{k_1,k_2}^{[2]}(b_a).
\end{align}
In addition, one can evaluate
\begin{align}\label{eq:normalization}
W_{k,k+1}(b_a)&=\begin{cases}
1 & \quad r+1:\mbox{odd}\\
\omega^{(-1)^{k}B_{\frac{r+3}{2}}} & \quad r+1:\mbox{even}
\end{cases}
\end{align}

Let us now introduce the Y-functions as \cite{y-system,ito-shu}
\be
\ba
{\cal Y}_{2j}(b_a)&=\frac{W_{-j,j}(b_a)W_{-j-1,j+1}(b_a)}{W_{-j-1,-j}(b_a)W_{j,j+1}(b_a)},\\
{\cal Y}_{2j+1}(b_a)&=\left[\frac{W_{-j-1,j}(b_a)W_{-j-2,j+1}(b_a)}{W_{-j-2,-j-1}(b_a)W_{j,j+1}(b_a)}\right]^{[+1]},
\ea
\ee
where $j \in \IZ_{\ge 0}$. By using repeatedly the Pl\"ucker type relation
\begin{align}
W_{k_{1},k_{2}}^{[+2]}W_{k_{1},k_{2}}&=W_{k_{1}+1,k_{2}+1}W_{k_{1},k_{2}}=-W_{k_{1}+1,k_{2}}W_{k_{2}+1,k_{1}}-W_{k_{1}+1,k_{1}}W_{k_{2},k_{2}+1},
\end{align}
we find
\begin{align}\label{eq:Y-system-zb}
{\cal Y}_{s}^{[+1]}(b_a){\cal Y}_{s}^{[-1]}(b_a)&=\Big(1+{\cal Y}_{s-1}(b_a)\Big)\big(1+{\cal Y}_{s+1}(b_a)\Big).
\end{align}
By definition, ${\cal Y}_0=0$. Note that there are $r+3$ sectors on the $z$-complex plane. Since 
$y_{j+r+3}(z)$ and $y_j( \re^{-2\pi \ri}z)$ are decaying solutions in the same sector, $y_{j+r+3}(z)\propto y_j( \re^{-2\pi \ri}z)$. We then obtain 
$y_{j+r+3}(z)\propto y_j( \re^{-2\pi \ri}z)=y_j(z)$, because the monodromy around the origin is trivial. We conclude that ${\cal Y}_{r+1}=0$. Therefore, we have $r$ Y-functions, which lead to a Y-system of the $A_r$-type. In the special case where $b_{1}=\cdots=b_{r-1}=0$, (\ref{eq:Y-system-zb}) reproduces the Y-system in the usual ODE/IM correspondence.

The system (\ref{eq:Y-system-zb}) is a generalization of the Y-system in \cite{dt-stokes, ito-shu}.
However, the parameters $b_a$ appearing on the l.h.s. are different from the ones in the r.h.s., and the Y-system is relating the Y-functions at 
different points of moduli space. To write down a closed TBA system, one should consider $b_a$ as independent parameters like in the derivation done in 
\cite{masoero}. However, this procedure will be rather complicated in general.
To avoid this problem and find a closed system, we introduce an spectral parameter $\zeta$, and we rescale the variables in (\ref{eq:Schrodinger-rescaled}) as 
\begin{align}\label{eq:rescale-su->zb}
q=\zeta^{\frac{2}{r+3}}z,\quad u_{a}&=-\zeta^{\frac{2(a+1)}{r+3}}b_{a},\qquad a=1,2,\cdots,r.
\end{align}
In the new variables, the Schr\"odinger equation (\ref{eq:Schrodinger-rescaled}) can be written as
\begin{align}\label{eq:Schrodinger-xu}
\left(-\zeta^2\partial_{q}^{2}+q^{r+1}-\sum_{a=1}^{r}u_{a}q^{r-a}\right)\hat{\psi}(q,u_a,\zeta)=0.
\end{align}
Therefore, the spectral parameter $\zeta$ plays the r\^ole of $\hbar$. 
We can now write the solution $y(z,b_a)$ in terms of $q$ and $u_a$:
\be
\hat{y}(q,u_{a},\zeta)=y(z,b_{a})=y(\zeta^{-\frac{2}{r+3}}q,-\zeta^{-\frac{2(a+1)}{r+3}}u_{a}).
\ee
The Symanzik rotation (\ref{sym-rotation}) can be written as a rotation of $\zeta$:
\begin{align}
(\omega^{-k}z,\omega^{-(a+1)k}b_{a})=\left((\re^{\ri\pi k}\zeta)^{-\frac{2}{r+3}}q,-(\re^{\ri\pi k}\zeta)^{-\frac{2(a+1)}{r+3}}u_{a}\right).
\end{align}
The solutions $y_k(z,b_a)$ can then be obtained from $\hat{y}(q, u_a,\zeta)$ by using
\be
\hat{y}_{k}(q,u_a,\zeta)=\omega^{\frac{k}{2}}\hat{y}(q,u_{a},\re^{\ri\pi k}\zeta).
\ee
Note that the solution does not go back to itself by rotating $\re^{2\pi \ri}\zeta$. Let us now introduce the Wronskian in terms of $\hat{y}_j$:
\be
\ba
\hat{W}_{k_1,k_2}(\zeta,u_a)&=\zeta^{\frac{2}{r+3}}\Big(\hat{y}_{k_1}(q,u_a,\zeta) \partial_q\hat{y}_{k_2}(q,u_a,\zeta)-\hat{y}_{k_2}(q,u_a,\zeta) \partial_q\hat{y}_{k_1}(q,u_a,\zeta)\Big)\\
&=
W_{k_1,k_2}(b_a).
\ea
\ee
We then introduce the Y functions, depending on the spectral parameter $\zeta$, as
\be\label{eq:new-Y}
Y_{2j}(\zeta,u_a)=\frac{\hat{W}_{-j,j}\hat{W}_{-j-1,j+1}}{\hat{W}_{-j-1,-j}\hat{W}_{j,j+1}}(\zeta,u_a),~~Y_{2j+1}(\re^{-\frac{\pi \ri}{2}}\zeta,u_a)=\frac{\hat{W}_{-j-1,j}\hat{W}_{-j-2,j+1}}{\hat{W}_{-j-2,-j-1}\hat{W}_{j,j+1}}(\zeta,u_a).
\ee
They satisfy
\begin{align}\label{eq:Y-system-xu}
Y_{s}(\zeta \re^{\frac{\pi \ri}{2}}, u_a)Y_{s}(\zeta \re^{-\frac{\pi \ri}{2}}, u_a)&=\Big(1+Y_{s+1}(\zeta, u_a)\Big)\big(1+Y_{s-1}(\zeta, u_a)\Big),
\end{align}
where $Y_0(\zeta,u_a)=Y_{r+1}(\zeta,u_a)=0$ (hereafter, we will omit the argument $u_a$ in the Y-functions.) The Y-systems 
(\ref{eq:Y-system-zb}) and (\ref{eq:Y-system-xu}) are essentially the same, but with different spectral parameters. 
Y-system (\ref{eq:Y-system-xu}) also appears in the study of minimal area surfaces in AdS$_3$ \cite{y-system,hito}.

Let us now calculate the small $\zeta$ behavior 
of the Y-functions (\ref{eq:new-Y}). The WKB expansion of $\hat{y}_k(q,u_a,\zeta)$ for small $\zeta$ is obtained  as
\be\label{eq:y-exact-WKB}
\hat{y}_k(q,u_a,\zeta)\sim (-1)^{\frac{k}{2}}c(\zeta)\exp\left(-\frac{\ri {\delta}_k}{\zeta}\int^{q}_{q^{(k)}}P(q')\rd q' \right). 
\ee
In this equation, $c(\zeta)={1\over \sqrt{2\ri}}\zeta^{\frac{r+1}{2(r+3)}}$, $P(q)$ was defined in (\ref{yyp}), $q^{(k)}$ is a point in the sector $\CS_k$, 
and ${\delta}_k=\pm (-1)^k$ is a sign factor, where $\pm$ depends on the sheet of the Riemann surface where $q$ lives. 
Using this WKB form, we can evaluate the Wronskian $\hat{W}_{k_1,k_2}(\zeta)$, which only depends on the initial points $q^{(k_1)}$ and $q^{(k_2)}$. Substituting this 
evaluation into the Y-functions (\ref{eq:new-Y}), we find that, as $\zeta\to 0$, the Y-functions behave as
\be
\ba
\log{Y}_{2k+1}(\zeta,u_{a})&\sim-\frac{1}{\zeta}\oint_{\gamma_{r-2k}}p(q)\rd q=-\frac{m_{r-2k}}{\zeta}, \\
\log{Y}_{2k}(\zeta,u_{a})& \sim-\frac{\ri}{\zeta}\oint_{\gamma_{r+1-2k}}p(q)\rd q=-\frac{m_{r+1-2k},}{\zeta}, 
\ea
\ee
which are valid in the region $|\mbox{arg}(\zeta)|<\pi$. Note that, with the definition (\ref{eq:new-Y}), the $Y_k$ function corresponds to the cycle that we labelled as $r+1-k$. From now 
on we will therefore set $Y_k \rightarrow Y_{r+1-k}$, which preserves the form of the Y-system (\ref{eq:Y-system-xu}). 

It is now a standard exercise to convert the Y-system (\ref{eq:Y-system-xu}) into a set of TBA equations, as done in \cite{y-system}. First we introduce an analytic function 
in $\left|\arg(\zeta) 
\right|\leq\frac{\pi}{2}$:
\be
\ell_a(\zeta):=\log Y_a(\zeta)+\frac{m_a}{\zeta}.
\ee
The Y-system then leads to
\begin{align}\label{eq:Y-system->ell}
\ell_{a}(\re^{\frac{\pi \ri}{2}}\zeta)+\ell_{a}(\re^{-\frac{\pi \ri}{2}}\zeta)=\log\Big((1+Y_{a+1}(\zeta))(1+Y_{a-1}(\zeta))\Big).
\end{align}
Let us set 
\be
\label{eps-def-Y}
\zeta=\re^{-\theta}, \qquad Y_a(\zeta)=\re^{-\epsilon_a(\theta)}
\ee
and let us introduce the kernel 
\be
K(\theta)= \frac{1}{2\pi \cosh(\theta)}. 
\ee
By convoluting (\ref{eq:Y-system->ell}) with this kernel and  exploiting the analyticity of $\ell_a(e^{-\theta})$, we obtain again the TBA equations (\ref{gentba}). 

Let us note that the $\epsilon$ functions introduced in (\ref{eps-def-Y}), in terms of the Y-functions defined in (\ref{eq:new-Y}), are not manifestly equal to the $\epsilon$ functions 
introduced in (\ref{def-eps}), which are defined as resummed quantum periods. 
Of course, it follows from our derivations that they satisfy the same TBA system with the same asymptotics, 
therefore they are indeed equal. Their equality should also follow by a direct comparison of their definitions, along the lines of \cite{allegretti}.

\subsection{Effective central charge and PNP relations}
\label{cc-pnp}

The TBA equations (\ref{gentba}) have some interesting general properties. 
First of all, the functions $\epsilon_a(\theta)$ have a constant value as $\theta \rightarrow 
-\infty$, which we will denote by 
\be
\label{cons-values}
\epsilon_a^\star= \lim_{\theta\to -\infty} \epsilon_a(\theta). 
\ee
As a consequence of (\ref{eq:Y-system->ell}), the corresponding $Y_a^\star$ values satisfy the equations 
\be
\left(Y_a^\star\right)^2 = (1+Y^\star_{a-1})(1+Y^\star_{a+1}). 
\ee
The solution to these equations is (see e.g. \cite{ddt})
\be
\label{solya}
Y_a^\star= { \sin \left(\frac{\pi  a}{r+3}\right) \sin
   \left(\frac{\pi  (a+2)}{r+3}\right) \over \sin^2\left(\frac{\pi }{r+3}\right)}. 
   \ee
It is possible to calculate the ``effective central charge" associated to this TBA system in terms of (\ref{cons-values}), by using the identity \cite{zamo-TBA,kl-mel-1}
\be
\label{c-charge}
c_{{\rm eff}}= {6 \over \pi^2} \sum_{a=1}^{r} m_a \int_\IR \re^\theta L_a(\theta) \rd \theta =
 r+ {3 \over \pi^2} \sum_{a=1}^{r} \left( \epsilon_a^\star \log\left(1+ \re^{\epsilon_a^\star}\right)+ 2 \, {\rm Li}_2(-\re^{\epsilon_a^\star})\right). 
\ee
This can be evaluated in closed form and one finds, 
\be
\label{ccharge-exp}
c_{\rm eff}={r(r+1) \over r+3}. 
\ee
In particular, this quantity is independent of the moduli. We will now relate the 
effective central charge to the so-called 
PNP relations, which provide a non-trivial 
and useful relation between the quantum periods. To simplify our discussion, 
we will assume that $r=2g$ is even. In this case, one can construct the following symplectic basis of cycles, 
\be
A_j=\gamma_{2j-1}, \qquad B_j= \sum_{k=j}^g (-1)^{k-j} \gamma_{2k}, 
\ee
where $\gamma_a$, $a=1,\cdots, 2g$ are the cycles defined in section \ref{sec-derivation}. We have chosen the cycles such that the 
intersection number of A and B cycles are $\langle A_j,A_k\rangle=0=\langle B_j,B_k\rangle$, $\langle A_j,B_k\rangle=\delta_{jk}$. Let us now 
introduce periods associated to this basis:
\be
\label{defnus}
\nu_j = {1\over 2 \pi} \oint_{A_j} P(q) \rd q, \qquad 
\nu_{D, j}=\ri   \oint_{B_j} P(q) \rd q. 
\ee
These are the quantum analogues of the $a$ and the $a_D$ variables in Seiberg--Witten theory. They have a power series expansion of the form, 
\be
\nu_j = \sum_{n \ge 0} \nu_j^{(n)} \hbar^{2n}, \qquad \nu_{D,j} = \sum_{n \ge 0} \nu_{D,j}^{(n)} \hbar^{2n},
\ee
We then define the quantum, or NS free energy, by 
\be
{\partial F^{\rm NS} \over \partial \nu_j}=\nu_{D, j}, \qquad j=1, \cdots, g. 
\ee
This is the analogue of the Nekrasov-Shatashvili (NS) free energy \cite{ns}, and it can be computed 
systematically as a formal power series of the form, 
\be
F^{\rm NS}(\nu)= \sum_{n \ge 0} F_n^{\rm NS}(\nu) \hbar^{2n}. 
\ee
Its classical limit, $F_0^{\rm NS}= F_0$, is the analogue of the Seiberg--Witten prepotential \cite{sw}. 
We will now introduce an additional quantum period which can be used to formulate the PNP relations.  
First of all we define the so-called ``projective coordinates" $X_j$, $j=0,1, \cdots, g$, as 
\be
X_0= {1\over \hbar}, \qquad \nu_j={X_j\over X_0}, \qquad j=1, \cdots, g. 
\ee
The quantum free energy can be promoted to a function of the projective coordinates as
\be
\CF(X, X_0)= \sum_{n \ge 0} X_0^{2-2n} F_n ^{\rm NS}( \nu). 
\ee
From this we construct the quantum period
\be
\CJ =-{1\over X_0} {\partial \CF \over \partial X_0}, 
\ee
which is a formal power series of the form 
\be
\CJ= \sum_{n \ge 0} \CJ_n \hbar^{2n}, 
\ee
and we have 
\be
\CJ_n = 2(n-1) F_n(\nu) + \sum_{j=1}^g \nu_j {\partial F_n ^{\rm NS} \over \partial \nu_j}, \qquad  n\ge 0. 
\ee
We note that the classical limit of $\CJ$ is given by 
\be
\CJ_0= -2F^{\rm NS}_0 + \sum_{j=1}^g \nu_j  {\partial F_0 \over \partial \nu_j}.  
\ee
In the case of Seiberg--Witten curves, $\CJ_0$ is a simple function of the moduli, in virtue of the so-called Matone relation \cite{matone} and its 
generalizations \cite{eguchi-yang,sty}. Let us now compute the next-to-leading order. From the definition of the quantum free energy, one has 
\be
{\partial F_1^{\rm NS} \over \partial \nu_j}= \nu_{D, j} - \sum_{k=1}^g \nu_k {\partial^2 F_0 \over \partial \nu_j \partial \nu_k}. 
\ee
It follows that
\be
\CJ_1= \sum_{j=1}^g \left( \nu^{(0)}_j \nu_{D,j}^{(1)}- \nu^{(1)}_j \nu_{D,j}^{(0)} \right). 
\ee
Up to a normalization, this is precisely the effective central charge associated to the TBA system, so one can evaluate
\be
\label{cj1}
\CJ_1=-{c_{\rm eff}\over 12}. 
\ee
In particular, this quantity is independent of the moduli. 

In quantum-mechanical models based on a genus one curve, with two independent quantum periods, 
one can find a non-trivial relation between them \cite{alvarez-casares,alvarez,dunne-unsal,gorsky,
basar-dunne, cm-ha,bdu-quantum}. We will refer to this relation as the PNP relation. It can be formulated in terms of the formal power series $\CJ$ as
\be
\label{pnp}
\CJ= \alpha E + \beta \hbar^2, 
\ee
where $E$ is the energy and $\alpha,\beta$ are constants. The leading order of this relation is just the analogue of Matone's relation for the corresponding elliptic curve, while the next-to-leading order of the equation states that $\CJ_1$ is a constant. This is precisely the content of (\ref{cj1}), and one finds that
\be
\label{beta-c}
\beta= -{c_{\rm eff}\over 12}. 
\ee
Let us note that (\ref{pnp}) is the ``integrated" version of the PNP relation presented in \cite{cm-ha}. The version one 
typically finds in the literature (e.g. in \cite{dunne-unsal}) 
is obtained from (\ref{pnp}) by taking an additional derivative w.r.t. $E$, and it misses the additional information carried by the constant $\beta$. 

\subsection{Wall-crossing}
\label{sec-wc}

The derivations we have presented in section \ref{sec-derivation} are valid in the ``minimal" chamber of moduli space, in which the masses are real. 
However, one can analytically continue the resulting TBA equations as follows. Let us assume that $m_a$ are now complex numbers, and let us write them as 
\be
m_a=|m_a| \re^{\ri\phi_a}, \qquad a=1, \cdots, r. 
\ee
Let us also introduce the shifted functions 
\be
\label{eq:shifted-epsilon}
 \widetilde \epsilon_a(\theta)= \epsilon_a(\theta- \ri \phi_a), \qquad  \widetilde L_a(\theta)= L_a(\theta- \ri \phi_a). 
 \ee
With this definition, $\widetilde \epsilon_a(\theta)\sim |m_a|e^\theta$ at large $\theta$ along the positive real axis. 
Using the shifted functions (\ref{eq:shifted-epsilon}), we obtain 
the TBA system 
\be\label{eq:TBA-complex-m-01}
\widetilde \epsilon_{a}(\theta)=|m_a|e^{\theta}-\int_{\IR } \frac{\widetilde L_{a-1}(\theta')}{\cosh(\theta-\theta'-\ri\phi_{a}+\ri\phi_{a-1})}\frac{\rd\theta'}{2\pi}
-\int_{\IR }\frac{\widetilde L_{a+1}(\theta')}{\cosh(\theta-\theta'-\ri\phi_{a}+\ri\phi_{a+1})}\frac{\rd\theta'}{2\pi}.
\ee
%
We denote the kernels appearing in the integral equations by\footnote{Our convention is different from the one in \cite{y-system}.}
\be
K_{r,s}(\theta)={1\over 2 \pi}{1\over \cosh\left(\theta + \ri (\phi_s-\phi_r) \right)}. 
\ee
The convolution will be denoted by:
\be
(K \star f)(\theta) = \int_\IR K(\theta-\theta') f(\theta') \rd \theta'. 
\ee
Finally, the superscripts $\pm$ denote a shift of the argument by $\pm \ri \pi/2$:
\be
f^{\pm}(\theta)= f\left( \theta \pm {\pi \ri \over 2} \right). 
\ee
The TBA equations (\ref{eq:TBA-complex-m-01}) can now be written in a more compact form as
\be
\widetilde \epsilon_{a}= |m_a|e^{\theta}- K_{a,a-1} \star \widetilde L_{a-1}-K_{a,a+1} \star \widetilde L_{a+1}. 
\ee

Note that the kernel of (\ref{eq:TBA-complex-m-01}) has poles when 
\be
|\phi_{a}-\phi_{a\pm 1}|=\frac{\pi}{2}.
\ee
Therefore, the equations (\ref{eq:TBA-complex-m-01}) are only valid in the region 
\be
\label{bound}
|\phi_{a}-\phi_{a\pm 1}|<\frac{\pi}{2}.
\ee
 As different $\phi_{a}-\phi_{a\pm 1}$ cross the values $\pm \frac{\pi}{2}, \pm \frac{3\pi}{2},\cdots$, we should modify the TBA equations to pick 
 the contributions of the poles. This phenomenon is known as the {\it wall crossing} of the TBA equations, and it goes back to the 
 discussion in e.g. \cite{dt-excited}. From 
 the point of view of resurgent Quantum Mechanics, what happens is that, as we move in moduli space, 
 the structure of discontinuities (or equivalently, of the Stokes automorphisms) becomes more complicated. In particular, more cycles are involved. This was 
 presciently described in \cite{reshyper} in terms of the appearance (or disappearance) of what they 
 called {\it geodesic cycles}, which are nothing but the BPS trajectories studied in e.g. \cite{selfdual,shapere-vafa,gmn2}. 
 In the case of the TBA system considered in \cite{gmn}, the wall-crossing phenomenon is indeed due 
 to the change of BPS spectra in the underlying $\CN=2$ gauge theory, as one moves in moduli space. In this paper, our study of wall-crossing 
 will rely on the approach put forward in \cite{y-system} and further developed by J. Toledo in \cite{toledo, toledo-thesis}, based on the analytic continuation of the 
 TBA system. 
 
 Let us start by considering the simple case of $r=2$. Let us suppose that we move in moduli space from a situation in which (\ref{bound}) holds, to a situation
 in which
\be
\phi_2 -\phi_1>{\pi \over 2}. 
\ee
Wall-crossing occurs when $\phi_2-\phi_1=\pi/2$. When this happens the periods satisfy the condition 
\be
\label{cms}
{\rm Im}\left( {\Pi_{\gamma_1} \over \Pi_{\gamma_2}}\right)=0, 
\ee
which defines a codimension one locus in moduli space. In $\CN=2$ theories, this locus is called the curve of marginal stability. 
The analytic continuation of the TBA system implies picking the residues of the kernel, and one easily finds \cite{y-system}
\be
\label{dw-tba-c2}
\ba
\widetilde \epsilon_1(\theta)&= |m_1| \re^{\theta} - K_{1,2} \star \widetilde L_2 
- L_2 \left( \theta-\ri \phi_1 -{\ri \pi \over 2}+\ri \delta \right), \\
\widetilde \epsilon_2(\theta)&= |m_2| \re^{\theta} - K_{2,1}\star \widetilde  L_1-L_1 \left(\theta-\ri \phi_2 +{\ri \pi \over 2}-\ri \delta \right). 
\ea
\ee
\begin{figure}[tb]
\begin{center}
\resizebox{105mm}{!}{\includegraphics{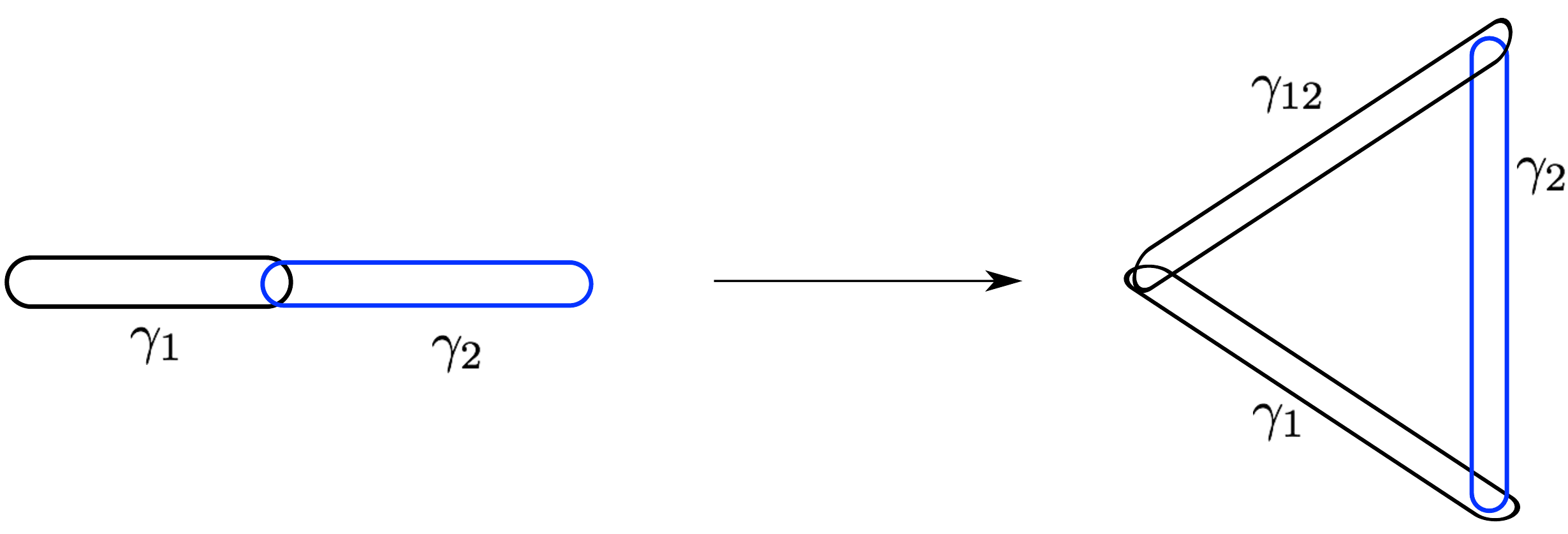}}
\end{center}
  \caption{As we move in moduli space away from the ``minimal" chamber, new cycles intervene in the TBA equations, and correspondingly, in the 
  discontinuity equations. The figure shows the situation for $r=2$. In the minimal chamber only the cycles $\gamma_1$, $\gamma_2$ are involved, but as they rotate in the complex plane, wall-crossing occurs and one has to consider as well the cycle $\gamma_{12}=\gamma_1+ \gamma_2$. 
}
\label{three-cycles}
\end{figure}
Due to the last terms in the r.h.s. of (\ref{dw-tba-c2}), the above system is not closed. We have to add the equations:
\be
\ba
\epsilon_1\left(\theta-\ri \phi_2+{\pi \ri \over 2}-\ri \delta\right) &= \ri  |m_1| \re^{\theta+ \ri \phi_1-\ri \phi_2-\ri \delta } 
- \int_\IR {L_2 \left(\theta'-\ri \phi_2 \right)
\over \ri \, \sinh(\theta-\theta'-\ri \delta )} {\rd \theta' \over 2 \pi}- \widetilde L_2\left( \theta\right), \\
\epsilon_2\left(\theta-\ri \phi_1-{\pi \ri \over 2} +\ri \delta  \right) &=-\ri  |m_2| \re^{\theta-\ri \phi_1+\ri \phi_2+\ri \delta} + 
\int_\IR {L_1 \left(\theta'-\ri \phi_1 \right)
\over \ri \sinh(\theta-\theta'+\ri \delta)} {\rd \theta' \over 2 \pi}- \widetilde L_1\left( \theta\right).
\ea
\ee
The resulting TBA system has in total four equations. It is possible to transform this system into another one 
involving {\it three} TBA equations for three new functions $ \epsilon^{\rm n}_{a}(\theta)=-\log Y_a^{{\rm n}}(\theta)$, 
where $a=1, 2, 12$. They are 
associated to the cycles $\gamma_1$, $\gamma_2$ and $\gamma_{12}= \gamma_1+\gamma_2$. 
The corresponding classical periods are $m_a$, $a=1,2$, like before, and
\be
m_{12}= m_1 -\ri m_2 =|m_{12} |\re^{\ri \phi_{12}}. 
\ee
These new functions are defined by, 
\be
\ba
Y_1^{\rm n}(\theta)&=\frac{Y_{1}(\theta)}{1+Y_{2}\left(\theta-\frac{\ri \pi}{2} \right)},\qquad 
\quad Y_{2}^{{\rm n}}(\theta)=\frac{Y_2 (\theta)}{1+Y_1\left(\theta+\frac{\ri\pi}{2} \right)},\\
\quad Y_{12}^{{\rm n}}(\theta)& =\frac{Y_{1}(\theta)Y_{2} \left(\theta-\frac{\ri\pi}{2} \right)}{1+Y_1(\theta)+Y_{2}\left(\theta-\frac{\ri\pi}{2}\right)}.
\ea
\ee
In order to write the new TBA system, we will remove the superscript ${\rm n}$ on the $\epsilon$ functions. The system reads,
\be
\label{new-cubic}
\ba
\widetilde \epsilon_1(\theta) &=  |m_1| \re^{\theta} -K_{1,2} \star \widetilde L_2 - K_{1,12}^+\star \widetilde L_{12}, \\
\widetilde \epsilon_2(\theta)&= |m_2| \re^{\theta} - K_{2,1} \star \widetilde L_1 - K_{2, 12} \star \widetilde L_{12},\\
\widetilde \epsilon_{12}(\theta) &= |m_{12}| \re^{\theta} - K_{12,1}^-\star  \widetilde L_1 - K_{12,2} \star \widetilde L_2.
\ea
\ee
This is the conformal limit of the system obtained in \cite{y-system,hito} in the study of minimal surfaces in AdS$_3$. 
In the context of $\CN=2$ gauge theory, the three TBA functions correspond to the three BPS states appearing after wall-crossing. A typical evolution in 
moduli space leading to the new system (\ref{new-cubic}) is shown in \figref{three-cycles}.

The pattern observed in this example can be extended to the general case \cite{toledo, toledo-thesis}. 
In addition to the minimal chamber we described in section \ref{sec-derivation}, there is a ``maximal" chamber 
involving $r(r+1)/2$ Y-functions, which are associated to cycles pairing {\it all} the turning points\footnote{It was already noted in \cite{reshyper} 
that there is such a chamber, involving $r(r+1)/2$ geodesic cycles. These results were later rediscovered in the study of 
BPS states of Argyres--Douglas (AD) theories \cite{shapere-vafa,gmn2}. This is not surprising, since AD theories are described by 
the same WKB curve as Quantum Mechanics with a polynomial potential (see e.g. \cite{ito-shu, gg-qm,gm-deformed}).}. The TBA equations for 
this ``maximal" chamber, in the case of minimal surfaces in AdS$_3$, have been obtained in \cite{toledo, toledo-thesis}, and they can be determined by a graphic procedure. The TBA equations for the WKB periods are simply the conformal limit of the equations in \cite{toledo, toledo-thesis}. There are 
also intermediate chambers where the number of Y-functions lies in between the minimal number $r$ and the maximal number $r(r+1)/2$. 
One can obtain the TBA equations in each intermediate chamber 
by carefully picking residues and redefining the Y-functions, as in the example with $r=2$ that we have described.

\begin{figure}[tb]
\begin{center}
\begin{tabular}{cc}
\resizebox{65mm}{!}{\includegraphics{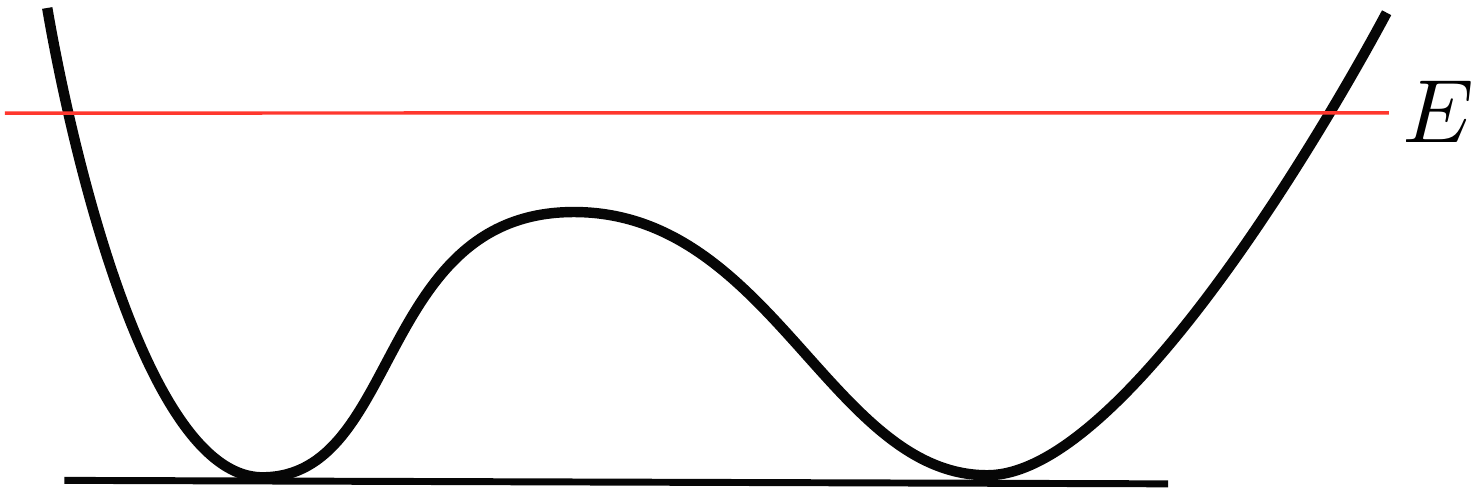}}
\hspace{10mm}
&
\resizebox{65mm}{!}{\includegraphics{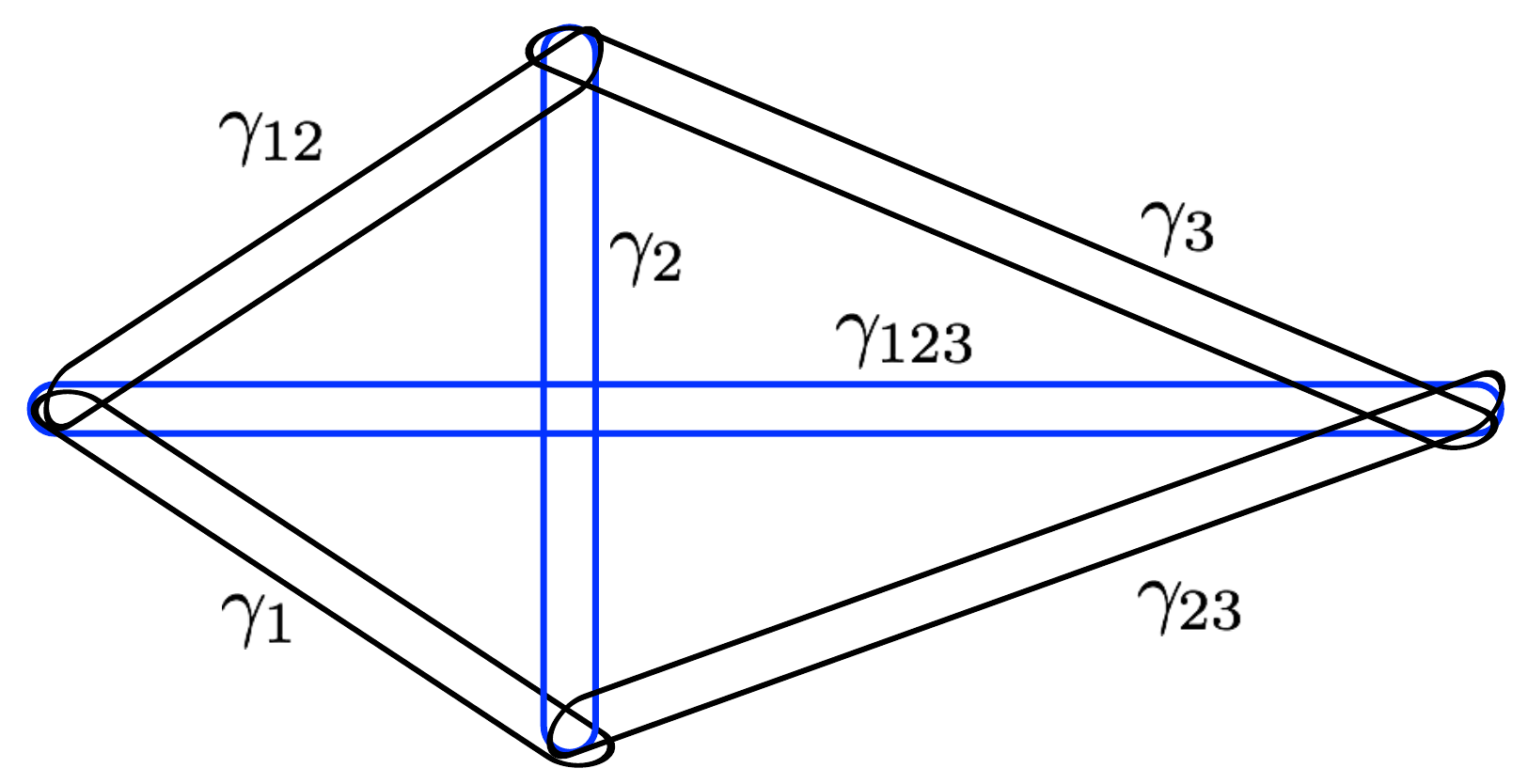}}
\end{tabular}
\end{center}
  \caption{In the case $r=3$, the ``maximal" chamber involves six TBA functions. This is the chamber relevant for a quartic potential in which the energy is above the potential barrier, as shown on the left. The TBA functions are associated to the cycles shown on the right. 
}
\label{new-cycles-quartic}
\end{figure}

Let us write down the resulting equations in the next case, $r=3$. 
The six TBA functions appearing after wall-crossing in the maximal chamber are associated to 
the cycles $\gamma_a$, $a=1,2,3$ appearing already in the minimal chamber, and to the new cycles (see \figref{new-cycles-quartic})
\be
\gamma_{12}= \gamma_1+ \gamma_2, \qquad \gamma_{23}= \gamma_2 +\gamma_3, \qquad \gamma_{123}= \gamma_1 + \gamma_2+ \gamma_3. 
\ee
The corresponding masses are 
\be
m_{12}= m_1 - \ri m_2, \qquad m_{23}= m_3 -\ri m_2, \qquad m_{123}= m_1 +m_3 -\ri m_2, 
\ee
and we will denote as usual 
\be
\phi_a= {\rm arg}(m_a), \qquad a=1,2,3,12, 23,123. 
\ee
The TBA system reads in this case, 
\be
\label{6tba}
\ba
\widetilde{\epsilon}_{1}&=|m_{1}|\re^{\theta}-K_{1,2}\star\widetilde{L}_{2}-K_{1,12}^{+}\star\widetilde{L}_{12}-K_{1,23}^{+}\star\widetilde{L}_{23}-K_{1,123}^{+}\star\widetilde{L}_{123},\\\widetilde{\epsilon}_{2}&=|m_{2}|\re^{\theta}-K_{2,1}\star\widetilde{L}_{1}-K_{2,3}\star\widetilde{L}_{3}-K_{2,12}\star\widetilde{L}_{12}-K_{2,23}\star\widetilde{L}_{23}-2K_{2,123}\star\widetilde{L}_{123},\\\widetilde{\epsilon}_{3}&=|m_{3}|\re^{\theta}-K_{3,2}\star\widetilde{L}_{2}-K_{3,12}^{+}\star\widetilde{L}_{12}-K_{3,23}^{+}\star\widetilde{L}_{23}-K_{3,123}^{+}\star\widetilde{L}_{123},\\\widetilde{\epsilon}_{12}&=|m_{12}|\re^{\theta}-K_{12,1}^{-}\star\widetilde{L}_{1}-K_{12,2}\star\widetilde{L}_{2}-K_{12,3}^{-}\star\widetilde{L}_{3}-K_{12,123}^{-}\star\widetilde{L}_{123},\\\widetilde{\epsilon}_{23}&=|m_{23}|\re^{\theta}-K_{23,1}^{-}\star\widetilde{L}_{1}-K_{23,2}\star\widetilde{L}_{12}-K_{23,3}^{-}\star\widetilde{L}_{3}-K_{23,123}^{-}\star\widetilde{L}_{123},\\\widetilde{\epsilon}_{123}&=|m_{123}|\re^{\theta}-K_{123,1}^{-}\star\widetilde{L}_{1}-2K_{123,2}\star\widetilde{L}_{2}-K_{123,3}^{-}\star\widetilde{L}_{3}-K_{123,12}^{+}\star\widetilde{L}_{12}
-K_{123,23}^{+}\star\widetilde{L}_{23}.
\ea
\ee
We will use a simplified version of this TBA system in section \ref{sec-quartic}. 

As we mentioned before, the discontinuities of the WKB periods in the maximal chamber are more complicated than in the minimal chamber (they involve 
more WKB periods). However, instead of deriving them from first principles as in \cite{voros-quartic}, it is much simpler to deduce them from the TBA equations obtained after wall-crossing.  
 
 The monic potentials appearing in the ODE/IM correspondence correspond to a point in moduli space with enhanced $\IZ_{r+1}$ symmetry which lies 
 in the maximal chamber. The resulting TBA system simplifies, due to this symmetry, and it coincides with the $A_r$ system considered in \cite{dt}. Therefore,
  the TBA system of the standard ODE/IM correspondence is a particular case of the 
 TBA system presented here, after wall-crossing. We will illustrate this in detail in the next sections, in the cases $r=2$ and $r=3$ (a closely related observation in the case $r=3$ has been 
 already made in \cite{toledo}). 
  
\sectiono{Example 1: the cubic potential}
\label{sec-cubic}

\begin{figure}[tb]
\begin{center}
\resizebox{65mm}{!}{\includegraphics{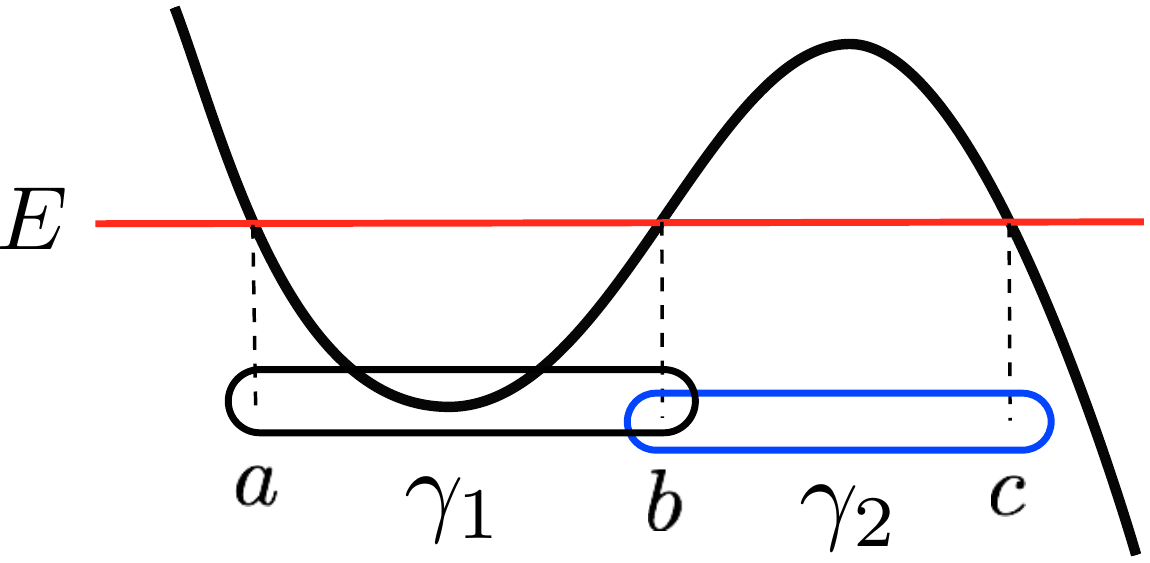}}
\end{center}
  \caption{The cubic oscillator. 
}
\label{cub-fig}
\end{figure}

\subsection{General aspects} 
Let us now illustrate the use of the TBA equations by applying them to the cubic oscillator, which is one of the most studied problems in Quantum Mechanics. 
The potential is taken to be, 
\be
\label{cubic-pot}
V(x)= {\kappa x^2\over 2}-  x^3. 
\ee
There are two WKB periods in this theory, associated to the two cycles 
shown in \figref{cub-fig}. The turning points will be denoted by $a, b, c$. The ``masses" introduced in (\ref{masses}) 
can be calculated explicitly in terms of elliptic integrals \cite{cm-ha}, and one has\footnote{We have followed the standard notation for elliptic integrals, which is different from the one used in e.g. {\it Mathematica}.}
\be
\ba
m_1&= {4 {\sqrt{2}} } { (b-c)^2 {\sqrt{a-c}} \over 15 k^4} \left( (k')^2 (k^2-2) K(k) + 2 (k^4 + (k')^2) E(k) \right),\\
m_2 &= 4\sqrt{2} \frac{(a-b)^2 (b-c)}{\sqrt{a-c}} \frac{2 (1+k^2 (k^2-1)) E(k') -k^2(1+k^2)K(k') }{15 k^2(k^2-1)^2},
\ea
\ee
where the elliptic modulus $k$ and its complementary $k'$ are given by 
\be
\label{kmod-cub}
k^2={b-c \over a-c}, \qquad (k')^2= 1- k^2 ={a-b \over a-c}. 
\ee
The TBA equations for this theory in the ``minimal" chamber are particularly simple:
\be
\label{cubic-tba}
\ba
\epsilon_1(\theta)&= m_1 \re^\theta - \int_\IR {\log\left(1+ \re^{-\epsilon_2(\theta')}\right)
\over\cosh(\theta-\theta')} {\rd \theta' \over 2 \pi}, \\
\epsilon_2(\theta)&= m_2 \re^\theta - \int_\IR {\log\left(1+ \re^{-\epsilon_1(\theta')}\right)
\over\cosh(\theta-\theta')} {\rd \theta' \over 2 \pi}. 
\ea
\ee
As we explained in section \ref{sec-wc}, this system is also valid for complex values of the masses as 
long as (\ref{bound}) is respected\footnote{
The cubic example considered in \cite{oper} corresponds to a value of the masses in which $m_1=m_2$ and 
$\phi_{1,2}=\pi/4$. In this case, the TBA system (\ref{cubic-tba}) reduces to a single equation.}. 
The asymptotic values of $\epsilon_{1,2}(\theta)$ as $\theta\rightarrow -\infty$ are given by
\be
\epsilon_\star= \log\left( {{\sqrt{5}}-1 \over 2} \right). 
\ee
The effective central charge is given by
\be
c_{\rm eff}={6\over 5}. 
\ee
On the other hand, the PNP relation for this theory (\ref{pnp}) can be written as \cite{cm-ha}
\be
\CJ= -{2 E \over 15}- {\hbar^2 \over 10}, 
\ee
so that (\ref{beta-c}) is verified. 

As a first application of the TBA system, we can look at the perturbative WKB expansion of the periods. The higher 
order corrections can be computed by various techniques, as we recall in Appendix \ref{app-ha}. They are highly non-trivial functions of the moduli of the WKB curve. At the same time, they can be obtained from the solution of the TBA system by using (\ref{qc-m}). In the original 
ODE/IM correspondence of \cite{dt,ddt}, which corresponds to a very particular point in the moduli space of the WKB curve, the integrals appearing 
in (\ref{qc-m}) can be computed analytically by relating them to the conserved charges of the 
underlying CFT \cite{blz1,blz2, blz-excited}. In this more general case, we do not know how to perform the integrals 
analytically, but we can solve the TBA system numerically, compute the coefficients $m_{1,2}^{(n)}$, 
and verify (\ref{mcorr-wkb}). In Table \ref{tab:cw-A} we compare $m_{1, 2}^{(n)}$, $n=1,2,3$, calculated numerically from the TBA system, to 
the quantum periods $\Pi_{\gamma_{1,2}}^{(n)}$, calculated with the techniques in Appendix \ref{app-ha}, for fixed values of the moduli $\kappa$ and $E$. 
As it can be seen in the Table, the two calculations 
agree to order $10^{-13}$, and the agreement can be improved by a more precise numerical integration of the TBA system. 

\begin{table}[tb]
\begin{center}
  \begin{tabular}{ccccc}\hline
	$n$ &  $\Pi_{\gamma_1}^{(n)} $                 &  $m_1^{(n)} $             &  $\Pi_{\gamma_2}^{(n)}$      &  $m_2^{(n)} $         \\ \hline
	$1$  & $3. 657 475 832 642 6$  & $-3. 657 475 832 642 3$  & $-9. 193 962 022 850 8$  & $-9. 193 962 022 850 5$        \\
	$2$ &$948.794 486 716 986$   & $948. 794 486 716 980$  & $19 138. 831 730 304 5$   & $19 138. 831 730 304 7$    \\
	$3$ & $1368 408. 366 655 312 $ & $-1 368 408. 366 655 315 $  & $-2 280 904 64. 0 893 14$  & $-2 280 904 64. 0 893 20 $   	 \\ \hline
  \end{tabular}
  \caption{The WKB expansion of the quantum periods for the cubic oscillator with $E=1/200$ and $\kappa=1$. The numerical calculation in the TBA system is done by Fourier discretization with $2^{12}$ points and a cutoff $(-L,L)$ where $L=35$.}
  \vspace{3mm}\label{tab:cw-A}
\end{center}
\end{table}

\subsection{PT-symmetric cubic oscillators}

\begin{figure}[tb]
\begin{center}
\resizebox{75mm}{!}{\includegraphics{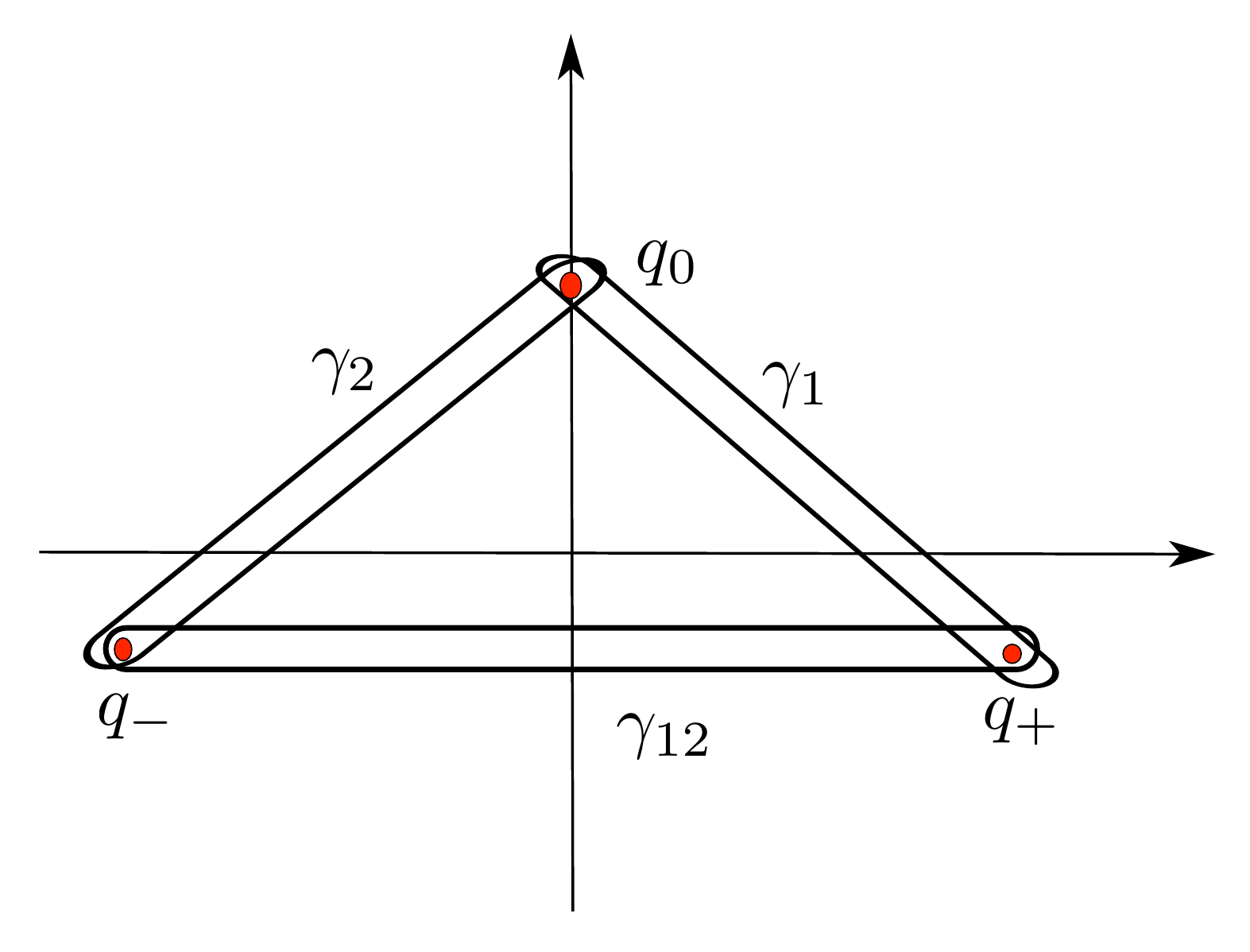}}
\end{center}
  \caption{The WKB cycles for the PT-symmetric cubic oscillator with Hamiltonian (\ref{pt-ham}). 
}
\label{ptcycles-fig}
\end{figure}

In order to illustrate the power of the TBA equations obtained in section 
\ref{sec-tba}, as applied to the cubic oscillator, we will analyze a concrete spectral problem 
which cannot be solved in the framework of the 
conventional ODE/IM correspondence. Let us consider the following Hamiltonian:
\be
\label{pt-ham}
\mH ={1\over 2}  \mm^2+ \ri \mq^3-\ri \lambda \mq. 
\ee
This Hamiltonian is PT-symmetric (see \cite{bender-review} for a review). 
When $\lambda\le 0$, it has a real, positive, discrete spectrum. When $\lambda=0$ we recover the PT-symmetric pure cubic oscillator. 
When $\lambda$ is positive and sufficiently large, PT symmetry is broken due to non-perturbative effects, 
and one finds an intricate pattern of level crossings studied in \cite{delabaere-trinh}.

We will focus here in the case in which $\lambda>0$. By rescaling $\hbar$ and the energy, it is 
easy to see that one can set $\lambda=1$. The potential 
\be
V(q)= \ri q^3 -\ri q 
\ee
has three turning points: one of them, $q_0$, is in the imaginary axis, while the other two, denoted by $q_\pm$, are 
in the fourth and the third quadrant of the complex plane, respectively (see \figref{ptcycles-fig}). 
All of them are non-trivial functions of the energy $E$. 
There are also three WKB cycles: the $\gamma_1$ cycle goes 
from $q_0$ to $q_+$, the $\gamma_2$ cycle goes from $q_-$ to $q_0$, and the cycle $\gamma_{12}=\gamma_1+\gamma_2$ goes from $q_-$ to $q_+$. The corresponding quantum periods are not independent, since they are related by 
\be
\Pi_{\gamma_2}=  \Pi_{\gamma_1}^*, \qquad \Pi_{\gamma_{12}}=\Pi_{\gamma_1}+\Pi_{\gamma_2}. 
\ee
In particular, the period $\Pi_{\gamma_{12}}$ is real. We will write  
\be
 \Pi_{\gamma_1}= {1\over 2}\Pi_{\rm p} - {\ri \over2}  \Pi_{\rm np}, 
 \ee
 where ${\rm p}$, ${\rm np}$ stand for ``perturbative" and ``non-perturbative." The perturbative period $\Pi_{\rm p}=\Pi_{\gamma_{12}} $ is 
the relevant one for the Bohr--Sommerfeld approximation, since the cycle $\gamma_{12}$ corresponds to 
classical, periodic {\it complex} trajectories between the turning points $q_-$ and $q_+$ and around them \cite{bb-pt, bender-review}. In this case, 
both $\Pi_{\rm p}$, $ \Pi_{\rm np}$ turn out to be Borel summable \cite{ddpham, dpham, delabaere-trinh}, and the all-orders, perturbative 
WKB approximation is given by 
  \be
  \label{pt-allwkb}
 s\left( \Pi_{\rm p} \right) (\hbar)= 2 \pi \hbar \left(n+{1\over 2} \right), \qquad n=0,1,2, \cdots
 \ee
However, this does {\it not} give the correct spectrum of the Hamiltonian (\ref{pt-ham}): 
as noted already in \cite{voros, bpv}, one has to take into account complex instantons\footnote{This is a quantum-mechanical example of how a Borel-resummable perturbative series 
fails to reproduce the exact results due to the appearance of complex instantons, 
as emphasized in \cite{gmz,cms} in a more general context.}. The correct EQC can be easily deduced from the analysis in 
 \cite{ddpham,alvarez-casares2, delabaere-trinh}, and it reads
 \be
 \label{eqc-pt}
 2 \cos\left( {1\over 2 \hbar}  s(\Pi_{\rm p})(\hbar) \right)+ \re^{- {1\over 2 \hbar} s(\Pi_{\rm np})(\hbar)}=0. 
 \ee
 The second term in the l.h.s. is an exponentially small correction which is crucial to obtain the exact results (in fact, this correction can become 
 exponentially enhanced when $s(\Pi_{\rm np})<0$, leading to PT-symmetry breaking \cite{delabaere-trinh,ben-ber}). When this correction is neglected, we recover (\ref{pt-allwkb}).
 
\begin{figure}[tb]
\begin{center}
\resizebox{70mm}{!}{\includegraphics{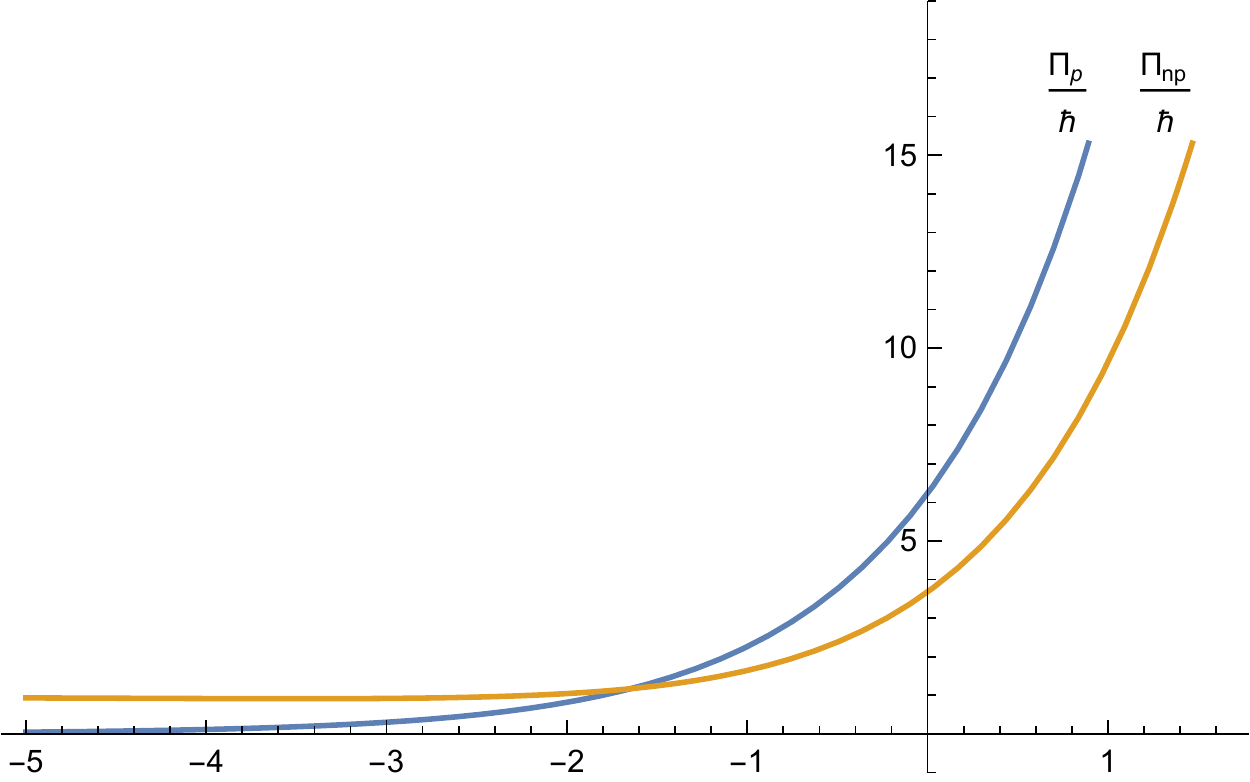}}
\hspace{3mm}
\end{center}
  \caption{Numerical calculation of the functions $s\left( \Pi_{\rm p}\right)(\hbar)/\hbar$ (top) and $s\left(\Pi_{\rm np}\right)(\hbar)/\hbar$ (bottom), as a function of $\theta=-\log(\hbar)$, by using the TBA equations. Here we set $E=1$.  
}
\label{pnp-plots}
\end{figure}

According to our results, the Borel-resummed quantum periods appearing in the EQC can be obtained from the TBA equations of Section \ref{sec-tba}. 
The PT-symmetric potential corresponds to a point in moduli space in the ``maximal" chamber. We then consider the 
three-term TBA equations in (\ref{new-cubic}). Let us denote
\be
\label{pt-angle}
\alpha= -{\rm arg}\left(\Pi_{\gamma_1} ^{(0)} \right). 
\ee
This angle is a non-trivial function of the energy $E$. The angles appearing in the kernels of (\ref{new-cubic}) are given by
\be
\label{pt-angles}
\phi_1= -\alpha, \qquad \phi_2= {\pi \over 2}+\alpha, \qquad \phi_{12}=0. 
\ee
We also have
\be
|m_1|= |m_2|= \left|  \Pi_{\gamma_{1}}^{(0)}  \right|, \qquad m_{12}=2\,  {\rm Re}\, \left( \Pi_{\gamma_{1}}^{(0)} \right). 
\ee
We can now solve the TBA system to obtain the three functions $ \widetilde \epsilon_a(\theta)$, $a=1,2, 12$. There is a conjugation symmetry
 \be
 \widetilde \epsilon_1(\theta)= \left( \widetilde \epsilon_2(\theta) \right)^*, \qquad \theta \in \IR. 
 \ee
As in (\ref{def-eps}), the relation between the quantum periods and the TBA functions is 
\be
\label{pis-eps}
-\ri \epsilon_1\left( \theta+{\ri \pi \over 2} \right) = {1\over \hbar} s\left(\Pi_{\gamma_1}\right)(\hbar), \qquad 
-\ri \epsilon_2 \left( \theta \right) = {1\over \hbar} s\left(\Pi_{\gamma_2}\right) (\hbar),
 \ee
where $\theta= -\log(\hbar)$. This implies
\be
\label{TBA-borel-2}
\ba
 {1\over \hbar} s\left(\Pi_{\gamma_1} \right)(\hbar) &=  {1\over \hbar} \Pi_{\gamma_1} ^{(0)} - \ri  \int_\IR {\widetilde L_2(\theta') \over \cosh\left(\theta- \theta'+\ri \alpha \right)} {\rd \theta' 
 \over 2 \pi}-\ri \int_\IR {\widetilde L_{12}(\theta') \over \cosh(\theta-\theta')} {\rd \theta'  \over 2 \pi},\\
  {1\over \hbar}s\left( \Pi_{\gamma_2} \right)(\hbar) &=  {1\over \hbar} \Pi_{\gamma_2} ^{(0)} +  \ri  \int_\IR {\widetilde L_1(\theta') \over \cosh\left(\theta- \theta'- \ri \alpha\right)} {\rd \theta' 
 \over 2 \pi}+\ri \int_\IR {\widetilde L_{12}(\theta') \over \cosh(\theta-\theta')} {\rd \theta'  \over 2 \pi}. 
 \ea
 \ee
By using these equations and the TBA system (\ref{new-cubic}), we can compute $s\left(\Pi_{\rm p} \right)$, $s\left(\Pi_{\rm np} \right)$ numerically, 
see \figref{pnp-plots}. 

One can now determine the Voros spectrum $x_n(E)$ for a fixed energy, by combining the EQC (\ref{eqc-pt}) with the TBA calculation of the resummed WKB periods. 
The accuracy of this method can be tested by calculating numerically the energies $E_n(x_n^{-1})$ corresponding to those values of the 
Planck constant. This is done by combining the standard Rayleigh--Ritz method with complex dilatation techniques 
\cite{bender-resonance}. We can then 
compare the numerical values obtained in this way to the expected relation (\ref{Exrelation}). The results are shown in Table \ref{tab:pt-def}, for 
$E=1$. The numerical approximation to the 
TBA system involves a discretization with $2^{18}$ points and a cutoff at $L=25$. As we can 
see, the numerical calculation of the spectrum by using the TBA system produces values which are correct with a precision of $10^{-10}$ (for the ground state) and of $10^{-12}$ for all the states with $n\ge 2$.  As in any semiclassical scheme, the approximation becomes better as the quantum number $n$ increases.

\begin{table}[tb]
\begin{center}
  \begin{tabular}{cccc}\hline
	$n$ & $x_n$                    &  $E_n (x_n^{-1})$             \\ \hline
	 $0$   & $0.\, 560 \, 405 \, 699 \,850 $  & $0. \,999 \,999 \,999 \,855$ \\
	 $1$   & $1.\,496 \,238\, 960 \,118 $  &$0. \,999 \,999 \,999 \,980$     \\
	 $2$   & $2.\,509 \, 053 \, 231 \, 083$ & $0. \,999 \,999 \,999 \,993$ \\  
	 $3$   & $3.\, 507 \,747 \, 393\, 173 $ & $0. \,999 \,999 \,999 \,997$ \\\hline
  \end{tabular}
  \caption{The first column shows the spectrum of values of $x_n$ for the Hamiltonian (\ref{pt-ham}), with $E=\lambda=1$. It is obtained by combining the EQC (\ref{eqc-pt}) with a numerical solution of the TBA system, involving a discretization with $2^{16}$ points and a cutoff at $L=25$. The second column shows the 
  values of the energy levels $E_n (x_n^{-1})$, computed by complex dilatation techniques.}
  \vspace{3mm}\label{tab:pt-def}
\end{center}
\end{table}

It is interesting to recover the results of the standard ODE/IM correspondence as a particular case of our more general results. 
Let us suppose that $\lambda=0$ in (\ref{pt-ham}), so that we obtain the pure PT-symmetric cubic oscillator
\be
\label{PT-pure}
\mH= {1\over 2}\mm^2 + \ri \mq^3. 
\ee
In this case we can set $E=1$ without loss of generality. The turning points are now located at the unit circle
\be
q_0= \ri, \qquad  q_+=\re^{-\pi \ri/6}, \qquad q_-=-\re^{\pi \ri/6}, 
\ee
and the classical periods can be calculated in closed form:
\be
\Pi_{\gamma_1}^{(0)}= {\sqrt{6 \pi} \Gamma(1/3) \over 3 \Gamma(11/6)}\re^{ -\ri \pi/3}, \qquad \Pi_{\gamma_2}^{(0)}= \left( \Pi_{\gamma_1}^{(0)} \right)^*. 
\ee
Higher order corrections can be computed by using the techniques of \cite{bpv}. We find, for example, 
\be
\ba
\Pi_{\gamma_1}^{(1)}&={\sqrt{6 \pi }  \Gamma \left(\frac{5}{3}\right) \over 8 \Gamma \left(\frac{1}{6}\right)} \re^{-2 \pi \ri/3}, \\
\Pi_{\gamma_1}^{(2)}&=0, \\
\Pi_{\gamma_1}^{(3)}&= {89  \sqrt{6 \pi} \Gamma\left(\frac{1}{3}\right) \over 120960 \Gamma\left(-\frac{19}{6}\right)} \re^{ -\pi \ri/3}.
\ea
\ee
The exact WKB theory of the Hamiltonian (\ref{PT-pure}) has a $\IZ_3$ symmetry which manifests itself in the TBA system. In particular, since 
\be
|m_1|= |m_2|= |m_{12}|, 
\ee
the $\widetilde \epsilon$ functions in (\ref{new-cubic}) 
are now all equal (and in particular they are real):
\be 
\label{all-equal}
\widetilde \epsilon_a(\theta)= \epsilon(\theta), \qquad a=1,2,3. 
\ee
The TBA system reduces to one TBA equation for the function $\epsilon(\theta)$ with periodicity $\epsilon(\theta)=\epsilon(\theta+\frac{5\pi i}{3})$. This equation can be easily written down, 
by using that the angle in (\ref{pt-angle}) is $\alpha=\pi/3$, and one finds
\be
\label{tba-cubic}
\epsilon(\theta) =  {\sqrt{6 \pi} \Gamma(1/3) \over 3 \Gamma(11/6)} \re^{\theta}+ \int_\IR \Phi(\theta-\theta') L(\theta') \rd \theta', 
\ee
where $L(\theta)=\log(1+ \re^{-\epsilon(\theta)})$, as usual, and 
\be
\Phi(\theta) = {\sqrt{3} \over \pi} {\sinh(2 \theta) \over \sinh(3 \theta)}. 
\ee
The TBA equation (\ref{tba-cubic}) appears in the analysis of the thermodynamics of 
the Lee--Yang model and it has been extensively studied in e.g. \cite{zamo-TBA,blz-excited}. It is the TBA equation that 
one would obtain for the $A_2$ system appearing in \cite{dt}. 
The relation between the quantum periods and the TBA functions is still given by (\ref{pis-eps}), and (\ref{TBA-borel-2}) 
further simplifies due to the $\IZ_3$ symmetry. We conclude that, when the potential is monic, the TBA system in the maximal 
chamber for $r=2$ (\ref{new-cubic}) reduces to the TBA equation of the conventional ODE/IM correspondence, as 
we mentioned in section \ref{sec-wc}. We have a similar result for the 
standard monic, cubic potential $V(q)=q^3$: the relation (\ref{all-equal}) still holds, where $\epsilon(\theta)$ satisfies (\ref{tba-cubic}), 
but the angles entering into the TBA system are in this case $\phi_1=-\phi_{12}= -\pi/6$, $\phi_2= \pi$.
\begin{table}[tb]
\begin{center}
  \begin{tabular}{ccccc}\hline
	$n$ &  $E_n^{\rm num} $                 &  $E_n^{\rm TBA}$             &  $E_n^{\rm p} $      &  $E_n^{\rm WKB}$         \\ \hline
	$0$  & $1.\, 156\, 267 \, 071 \, 988$  & $1.\, 156\, 267 \, 071 \, 988$  & $1.\, 134\, 513 \, 239 \, 424$  & $1.\, 094\, 269 \, 500 \, 533$        \\
	$1$ &$ 4.\, 109\,228 \,752 \,810 $   & $4.\, 109 \, 228 \,752 \,806$  & $4.\, 109 \, 367 \,351 \,095$   & $4.\, 089 \, 496 \,119 \,273 $    \\
	$2$ & $7.\, 562 \,273 \,854 \,979$ & $7.\,562 \, 273\, 854 \, 971 $  & $7.\,562 \, 273\, 170 \, 784 $  & $7.\,548 \, 980\, 437 \, 586  $   	 \\ \hline
  \end{tabular}
  \caption{The energy levels for the PT cubic oscillator for $\hbar={\sqrt{2}}$. The first column shows a numerical calculation with complex dilatation 
  techniques. The second column is a calculation by using the TBA system with $L=25$ and $n=2^{16}$ steps in the discretization. The third 
  column is obtained by using the all-orders WKB perturbative quantization condition (\ref{pt-allwkb}) and the TBA calculation of $\Pi_{\rm p}(\hbar)$. The final column is 
  the result of the Bohr--Sommerfeld approximation. }
  \vspace{3mm}\label{tab:pt-ens}
\end{center}
\end{table}

The EQC for (\ref{PT-pure}) is again given by (\ref{eqc-pt}). Note that, in this case, the energy $E$ and the Planck constant always 
appear in the combination 
\be
{E^{5/6} \over \hbar}
\ee
and the Voros spectrum can be easily translated into the usual spectrum. In Table \ref{tab:pt-ens} we compare different 
approaches to the calculation of the energy levels of this oscillator. We have chosen $\hbar={\sqrt{2}}$ to facilitate the comparison with results in the literature \cite{bender-review, bb-pt}\footnote{Note however that the some of the numerical results for the excited energy levels of (\ref{PT-pure}) presented in Table 1 
of \cite{bender-review} are slightly off.}. The first column is a numerical calculation of the energy levels, by combining 
standard Rayleigh--Ritz techniques with complex dilatation. The 
 second column uses the EQC (\ref{eqc-pt}) and the TBA calculation of the quantum periods. The third column is obtained by using the all-orders WKB perturbative quantization condition (\ref{pt-allwkb}), and the TBA calculation of $s\left(\Pi_{\rm p}\right)$. The final column gives the result of the Bohr--Sommerfeld quantization condition, which approximates the l.h.s. of (\ref{pt-allwkb}) by the classical period. 
 
\begin{figure}[tb]
\begin{center}
\begin{tabular}{cc}
\resizebox{55mm}{!}{\includegraphics{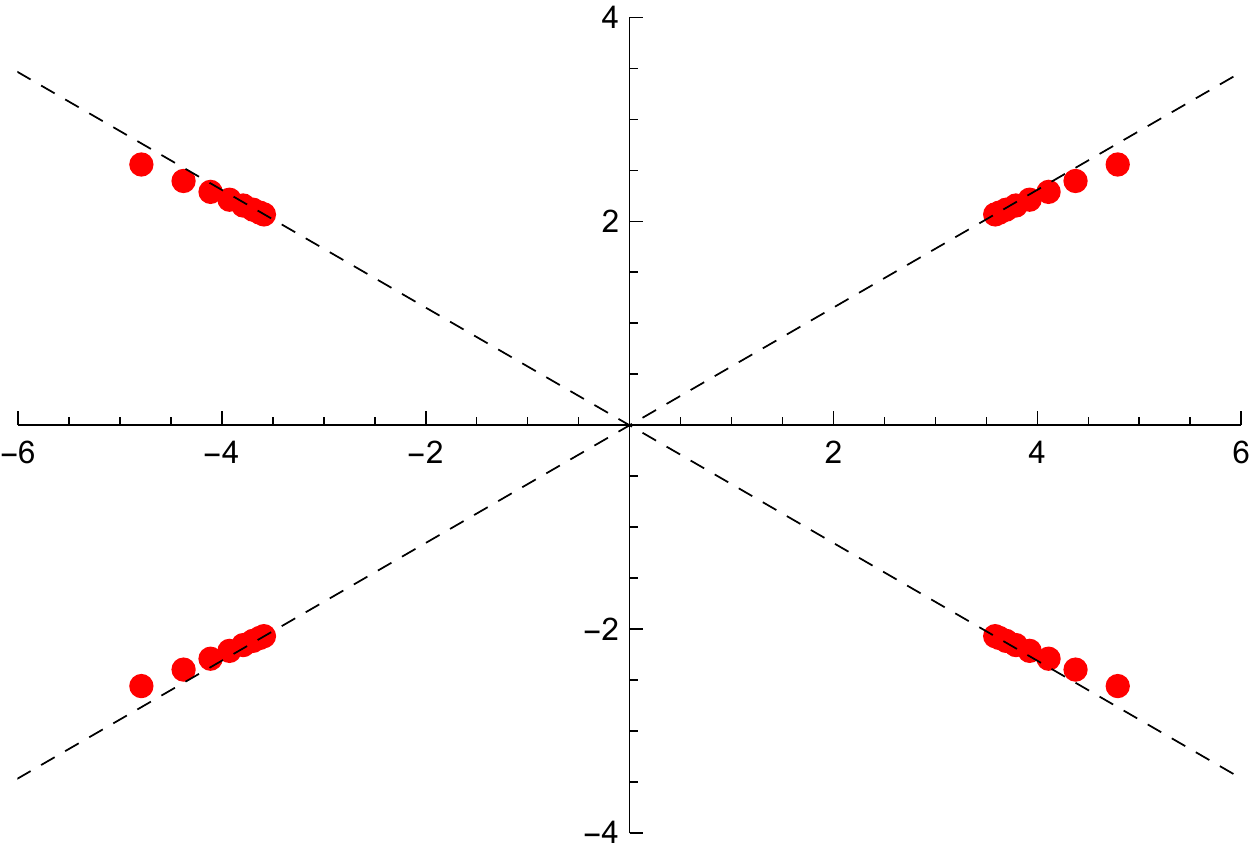}}
\hspace{10mm}
&
\resizebox{55mm}{!}{\includegraphics{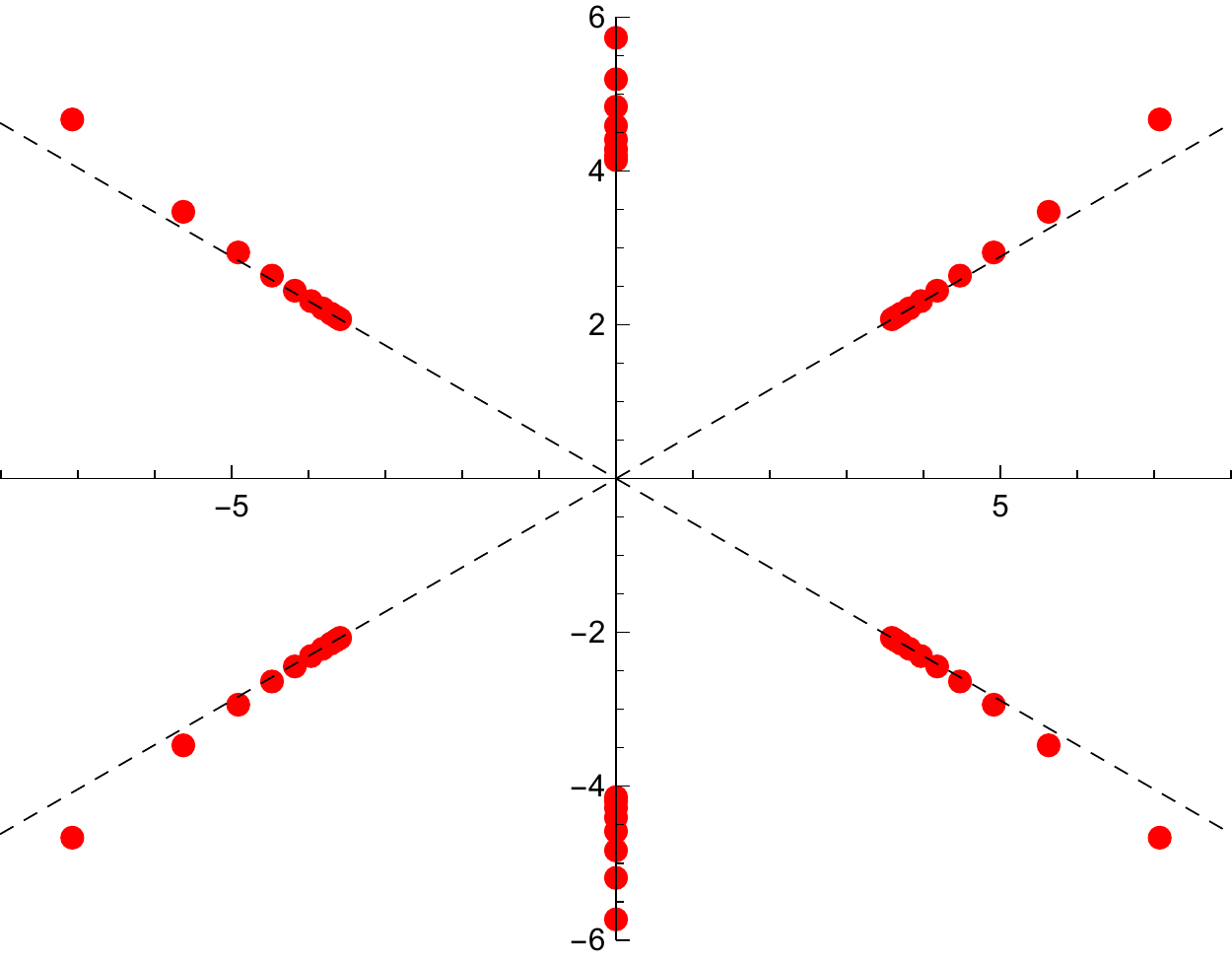}}
\end{tabular}
\end{center}
  \caption{The singularity structure of the Borel transforms of $\Pi_{\rm p}$ (left) and 
$\Pi_{\rm np}$ (right) for the PT-symmetric pure cubic oscillator (\ref{PT-pure}) can be approximated numerically by 
using Borel--Pad\'e techniques, and it reproduces the discontinuities of the TBA equations. The diagonal rays are at the angles $\pm \pi/6$. 
}
\label{borel-planes}
\end{figure}

As we have already noted, although the structure of discontinuities of the WKB periods in the maximal chamber is more complicated than the one in the 
minimal chamber, it can be determined directly from the TBA equations (\ref{new-cubic}). We would also like to emphasize that many ingredients of this structure 
can be determined empirically by looking at the singularities of the Borel transform, in agreement 
with the theory of resurgence. The PT-symmetric pure cubic oscillator (\ref{PT-pure}) provides a beautiful illustration 
of this fact. The discontinuities for $\Pi_{\rm p}/\hbar $ and $\Pi_{\rm np}/\hbar$ can be easily obtained from (\ref{new-cubic}). 
In the case of the perturbative period, they are located at the rays with angles $\pm \pi/6$. For the non-perturbative period, there are 
discontinuities at the rays with angles $\pm \pi/6$ and $\pm \pi/2$. A numerical calculation of the Borel transforms of $\Pi_{\rm p}$ and 
$\Pi_{\rm np}$ is shown in \figref{borel-planes}, by using standard Borel--Pad\'e techniques applied to $125$ terms of their formal power series. The location of 
the singularities reproduces precisely the location of the discontinuities obtained from the TBA system.

\sectiono{Example 2: the quartic potential}
\label{sec-quartic}

 \begin{figure}[tb]
\begin{center}
\resizebox{75mm}{!}{\includegraphics{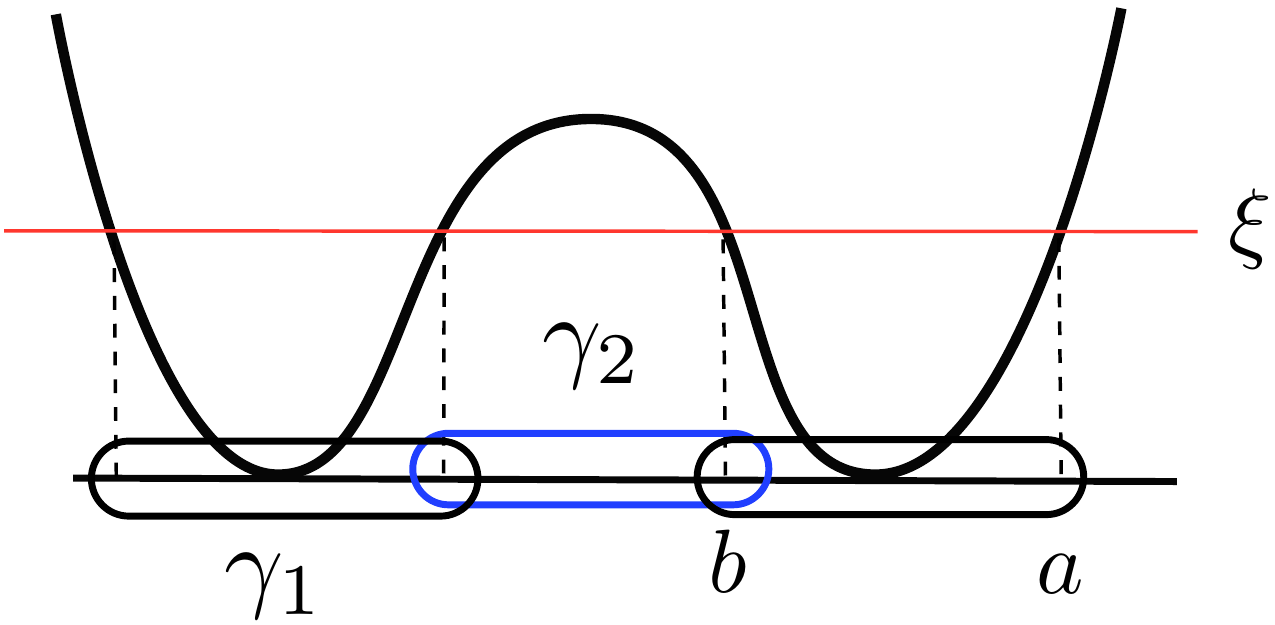}}
\end{center}
  \caption{The symmetric double-well potential. 
}
\label{dw-fig}
\end{figure}

\subsection{General aspects}

As another example of our general results, we consider Quantum Mechanics with a 
quartic potential. For simplicity, we will mostly focus on the case of a parity symmetric potential obeying $V(x)= V(-x)$. 
A convenient parametrization of such a potential is
\be
\label{dw-potential}
V(x)={1\over 32}-{\kappa x^2 \over 4} +{x^4\over 2}.
\ee
When $\kappa=1$, we have the double-well potential 
\be
V(x)= {1\over 32} \left( 4 x^2-1\right)^2, 
\ee
with the structure shown in \figref{dw-fig}. In this case, the turning points are real when the energy satisfies 
\be
0<E <{1\over 32}. 
\ee
When $\kappa<0$, we have a perturbed harmonic oscillator.   

Let us now consider the TBA equations governing the WKB periods of this potential. Like in the cubic case, we first consider the minimal chamber. 
Due to the parity symmetry, $\gamma_1= \gamma_3$, so the corresponding TBA functions appearing in (\ref{gentba}) are equal, 
\be
\epsilon_1=\epsilon_3, 
\ee
and the general equations (\ref{gentba}) for $r=3$ simplify to 
\be
\label{dw-tba}
\ba
\epsilon_1(\theta)&= m_1 \re^\theta - \int_\IR {\log\left(1+ \re^{-\epsilon_2(\theta')}\right)
\over\cosh(\theta-\theta')} {\rd \theta' \over 2 \pi}, \\
\epsilon_2(\theta)&= m_2 \re^\theta - 2\int_\IR {\log\left(1+ \re^{-\epsilon_1(\theta')}\right)
\over\cosh(\theta-\theta')} {\rd \theta' \over 2 \pi}. 
\ea
\ee
The classical periods can be computed explicitly in terms of elliptic integrals (see e.g. \cite{cm-ha}), and one finds, 
\be
\label{dw-expers}
\ba
m_1&=  {2 a ^3 \over 3  }\left[ (1+ k'^{2}) E(k) - 2 k'^{2} K(k) \right],\\
m_2&={4 a^3 \over 3 } \left[ (1+k'^{2}) E(k')- k^2 K(k') \right],
\ea
\ee
where the elliptic moduli are
\be
\label{k-dw}
k^2=1-{b^2 \over a^2}, \qquad \left(k'\right)^2=1-k^2. 
\ee
By solving the TBA system numerically, we can compute quantum corrections to the periods and compare them to the analytic 
results obtained in Appendix \ref{app-ha}. 
In particular, we can check again the identification (\ref{mcorr-wkb}) with high precision and to high orders. 

The limiting behavior of $\epsilon_1$, $\epsilon_2$ is easily found from (\ref{solya}) to be
\be
\epsilon_1^\star= -\log(2), \qquad \epsilon_2^\star =-\log(3),  
\ee
and the effective central charge is 
\be
\label{c-iden}
c_{\rm eff}=2. 
\ee
Although our general discussion of the PNP relation in section \ref{cc-pnp} was carried out for simplicity for odd potentials, 
it can be extended straightforwardly to the double-well potential. By using the period and dual period defined in e.g. \cite{cm-ha}, 
one finds that the PNP relation for this potential reads
\be
\CJ=-{E \over 3}- {\hbar^2 \over 6}, 
\ee
so that $\alpha=-1/6$, and (\ref{beta-c}) is again satisfied.

\begin{figure}[tb]
\begin{center}
\resizebox{75mm}{!}{\includegraphics{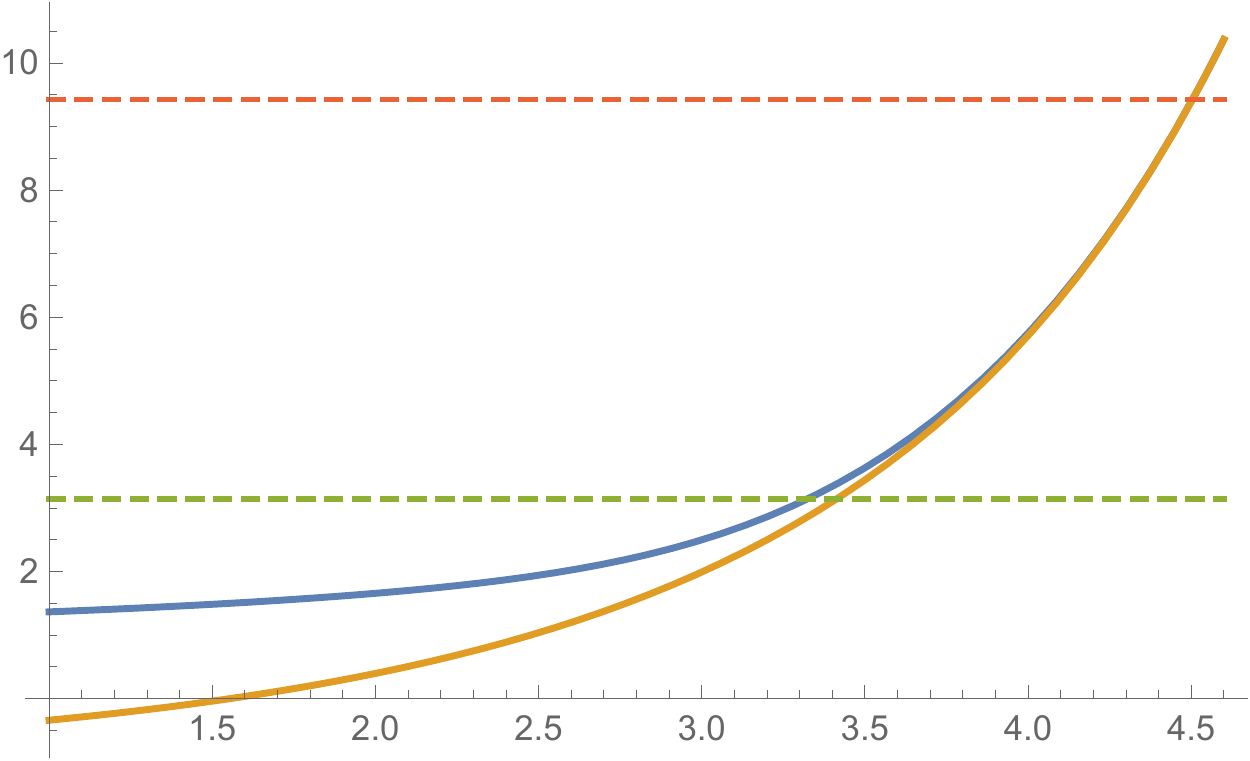}}
\end{center}
  \caption{The functions $\CJ_{\epsilon}(\hbar)$ for $\epsilon=-1$ (bottom) and $\epsilon=1$ (top) as a function 
  of $x=-\log \hbar$. The horizontal lines are $\pi$ and $3 \pi$, and intersect $\CJ_\epsilon(\hbar)$ 
  in four points which give the first four values of the spectrum of $\hbar$. 
}
\label{eqc-fig}
\end{figure}

\subsection{Quartic oscillators }

Let us now illustrate how to use the TBA system to compute the spectrum of various quartic oscillators. 

Let us first consider the double-well potential. When the energy of the system is below 
the top of the barrier, we are in the minimal chamber and we can use the TBA equations (\ref{dw-tba}) to compute 
the Borel resummation of the WKB periods. In this case, as it follows from the Delabaere--Pham formula, 
the quantum period $\Pi_{\gamma_1}$ is not Borel summable for real values of $\hbar$, and one has to specify an appropriate 
resummation. As it should be clear from the discussion in section \ref{sec-derivation}, the lack of Borel summability of $\Pi_{\gamma_1}$ 
is manifest in the TBA system: after shifting $\theta \rightarrow \theta + \pi \ri/2$ in $\epsilon_1(\theta)$, as prescribed by (\ref{def-eps}), we hit 
a pole in the kernel and the integral becomes singular. The two deformations of the integration contour that avoid this pole correspond to the two lateral Borel resummations, as indicated in (\ref{def-eps}) by the explicit introduction of a small $\delta$. In many cases, however, the most appropriate 
resummation is the so-called ``median" resummation \cite{dpham,ddpham} (see \cite{mmnp,as13} for other applications of  
the median resummation). We will denote this resummation by $s_{\rm med}\left(\Pi_{\gamma_1}\right)$, which in this particular example is just the 
average of the two lateral resummations above and below the Stokes line. In the TBA system, the median resummation is easily implemented: it corresponds to 
the principal value of the singular integral. 

To compute the resummed quantum periods, we first solve the TBA system 
(\ref{dw-tba}) to obtain the functions $\epsilon_{1,2}(\theta)$ along the real line. We then have, 
\be
\ba
{1\over \hbar} s_{\rm med}\left( \Pi_{\gamma_1}\right)(\hbar)&= m_1 \re^{\theta}+ {\rm P} \int_\IR {L_2(\theta') \over \sinh(\theta-\theta')} {\rd \theta' \over 2 \pi}, \\
{\ri \over \hbar} s\left(\Pi_{\gamma_2} \right) (\hbar)&=\epsilon_2(\theta), 
\ea
\ee
where as before we have to identify $\theta=-\log(\hbar)$. 

The EQC for this problem was famously conjectured by Zinn--Justin in \cite{zinn-justin} (it follows however by a simple analytic continuation from previous results 
by Voros \cite{voros} for the pure quartic oscillator). In the language of resurgence, Zinn-Justin's EQC has been derived and reformulated in \cite{ddpham}, and it can be 
written in the following way: 
\be
\cos\left({1\over \hbar}  s_{\rm med} \left(\Pi_{\gamma_1}(\hbar) \right) \right) +{1\over {\sqrt{1+ \re^{- \ri s\left( \Pi_{\gamma_2}(\hbar) \right)/\hbar} }}}=0. 
\ee
This can be solved more explicitly as follows, 
\be
\label{eqc-dw}
\CJ_{\epsilon}(\hbar)=  2 \pi \left( k+{1\over 2} \right),  \qquad k \in \IZ_{\ge 0}, 
\ee
where
\be
\label{j-eqc-dw}
\CJ_{\epsilon}(\hbar)= {1\over \hbar} s_{\rm med} \left(\Pi_{\gamma_1}\right)(\hbar) 
+\epsilon \tan^{-1}\left( \re^{- {\ri \over 2 \hbar} s\left( \Pi_{\gamma_2} \right)(\hbar) } \right). 
\ee
Each integer value of $k$ leads to two states of the spectrum, depending on the value of the parity $\epsilon=\pm 1$. In this way, the 
$n$-th state of the spectrum, where $n=0,1,2, \cdots$, corresponds to 
\be
\label{nmepsilon}
n=2k-{\epsilon-1 \over 2}. 
\ee
The median resummation $s_{\rm med} \left(\Pi_{\gamma_1}\right) $ can be regarded as the all-orders, perturbative WKB contribution, 
while the Borel resummed period $\ri s\left(\Pi_{\gamma_2}\right)$ gives a non-perturbative 
correction due to tunneling effects. 

Using the TBA equations, we have computed the functions 
$\CJ_{\pm}(\hbar)$ appearing in the l.h.s. of the exact quantization condition, as well as the Voros spectrum $x_n(E)$. 
In \figref{eqc-fig} we plot the functions $\CJ_{\epsilon}$ for $E=1/64$, as a function of $\theta$, as well as the horizontal lines at $\pi$ and $3 \pi$, 
which intersect $\CJ_{\pm}(\hbar)$ in four points which give the first four values of the spectrum of $\hbar$. For the numerical 
solution of the TBA system we have used a Fourier discretization in an interval $(-L,L)$ with $L=35$ and $N=2^{18}$ points. The principal value is computed as the limit
\be
\label{pv-pres}
{\rm P} 
\int_\IR {f(\theta') \over \sinh(\theta-\theta)} \rd \theta'= \lim_{\delta \to 0} \int_{\IR} {\sinh(\theta-\theta') \cos(\delta)  f(\theta') \over
 \sinh^2(\theta- \theta') \cos^2(\delta) + \cosh^2(\theta- \theta') \sin^2(\delta) } \rd \theta', 
\ee
which is obtained by considering the real part of 
\be
{1\over \sinh(\theta-\theta'\pm \ri \delta)}. 
\ee
Numerical approximations to the principal value are then obtained by picking a small value of $\delta$. In our numerical computation, we have used $\delta=10^{-15}$. 
The values of $x_n$ obtained in this way can be tested numerically by verifying the identity (\ref{Exrelation}), where the energy levels are those of the Hamiltonian
\be
\label{h-dw}
\mH (\hbar)={\mm^2 \over 2}-{\mq^2 \over 4} + {\mq^4 \over 2}+{1\over 32}. 
\ee
An example of such a computation, for $E=1/64$, is shown in Table \ref{tab:dw}. Although the precision obtained for the quartic potential is lower 
than for the cubic potential discussed in the previous section, the discrepancies are at worst of $10^{-8}$ (for the ground state).

\begin{table}[tb]
\begin{center}
  \begin{tabular}{cccc}\hline
	Level & $x_n$                    &  $E_n (x_n^{-1})$             \\ \hline
	 $0$   & $27.\, 792 \, 941\, 509$  & $0.\, 015\,  625\,  014 $                \\
	 $1$   & $30.\, 383\,  260\, 063$  &$0.\,015 \,625\, 011$     \\
	 $2$   & $90.\, 374\, 172\, 968$ & $0.\,015\, 625\, 000 $ \\  
	 $3$   & $90.\,  400\, 201\, 849$ & $0.\,015 \,625 \,000 $ \\ \hline
  \end{tabular}
  \caption{The first column shows the spectrum of values of $x_n$, as computed from the exact quantization condition (\ref{j-eqc-dw}) and the TBA equations. We used 
  a discretization with $2^{18}$ points and $L=35$. The second column 
  shows the values of $E_n (\hbar_n)$ for the Hamiltonian (\ref{h-dw}).}
  \vspace{3mm}\label{tab:dw}
\end{center}
\end{table}

\begin{figure}[tb]
\begin{center}
\resizebox{65mm}{!}{\includegraphics{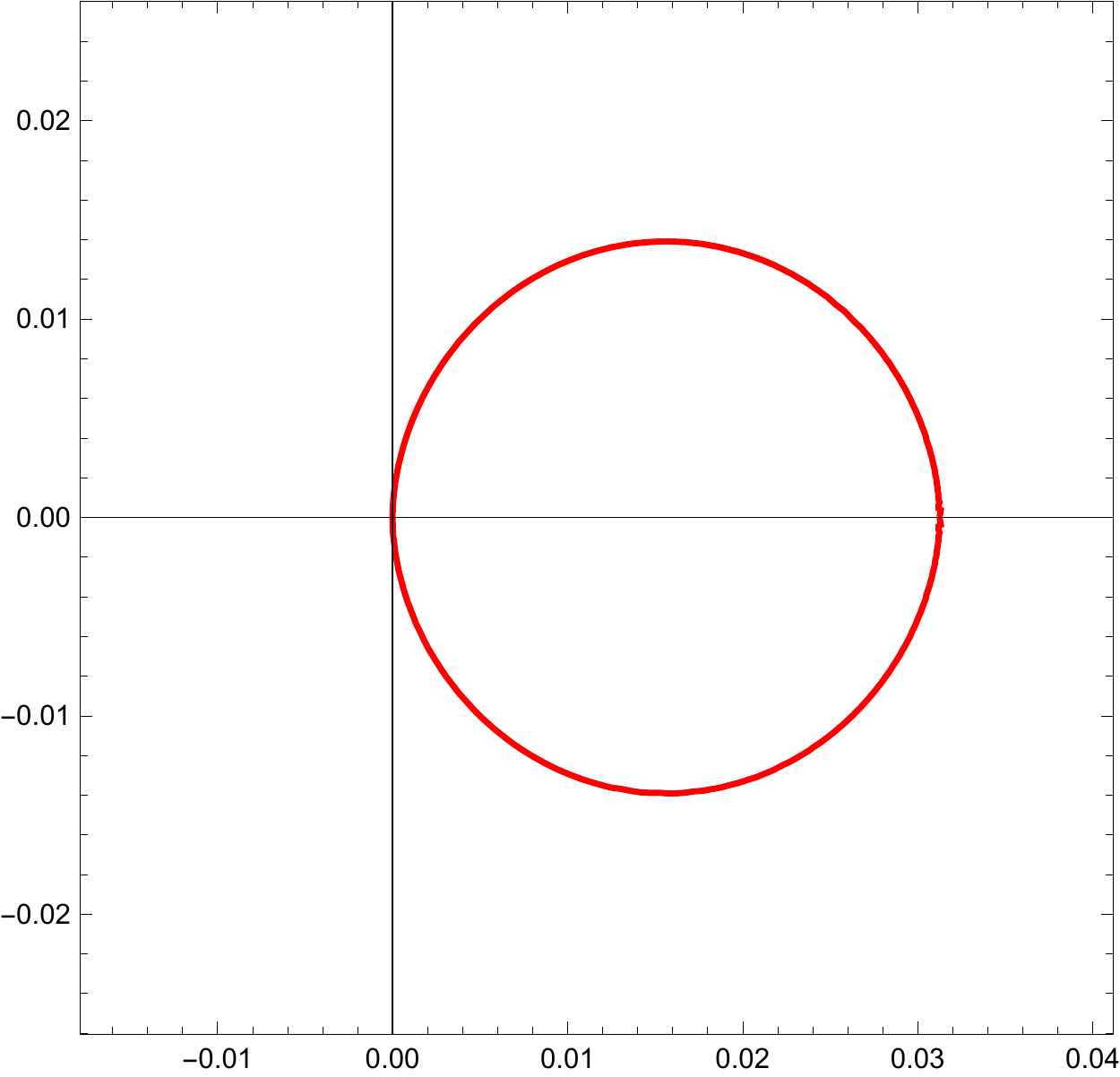}}
\end{center}
  \caption{The curve of marginal stability for the quartic oscillator with $\kappa=1$, in the complex energy plane. In the real axis, it stretches from $E=0$ to the top 
  of the barrier at $E=1/32$.
}
\label{cms-fig}
\end{figure}

In the previous example we considered a quartic potential in the minimal chamber: the double-well potential with an energy below the barrier. However, as soon as 
the energy increases above the barrier, we move into the maximal chamber of moduli space, as shown in \figref{new-cycles-quartic}. 
More generally, one can easily determine the counterpart of the curve of marginal stability for the double-well potential in the complex energy plane, defined 
by the condition (\ref{cms}). It is shown in \figref{cms-fig}. The minimal chamber, where (\ref{dw-tba}) holds, corresponds to the region inside the curve. 

Let us then focus on the maximal chamber. The corresponding TBA system was written down in (\ref{6tba}) and it involves generically six functions. 
However, due to the parity 
symmetry of the symmetric quartic potential, we have
\be
m_3=m_1, \qquad m_{12}= m_{23}, 
\ee
and we can set 
\be
\widetilde \epsilon_3=\widetilde \epsilon_1, \qquad \widetilde \epsilon_{23}=\widetilde \epsilon_{12}. 
\ee
The resulting TBA system involves only four functions, $\widetilde \epsilon_a$ with $a=1,2,12,123$, and it reads
\be
\label{6tba-red}
\ba
 \widetilde \epsilon_1 &=  |m_1| \re^{\theta} -K_{1,2} \star \widetilde L_2 - 2 K_{1,12}^+\star \widetilde L_{12}-K_{1,123}^+\star \widetilde L_{123}, \\
\widetilde \epsilon_2&=|m_2| \re^{\theta} -2 K_{2,1} \star \widetilde L_1 -2 K_{2, 12} \star \widetilde L_{12}-2K_{2, 123} \star \widetilde L_{123},\\
\widetilde \epsilon_{12}&=|m_{12}| \re^{\theta} -2  K^-_{12,1} \star \widetilde L_1 -K_{12,2} \star \widetilde L_{2}
-K^-_{12, 123} \star \widetilde L_{123},\\
 \widetilde \epsilon_{123}&=|m_{123}| \re^{\theta} -2 K^-_{123,1} \star \widetilde L_1 -2 K_{123,2} \star \widetilde L_{2}-2K^+_{123, 12} \star \widetilde L_{12}.
\ea
\ee

Let us now apply this TBA system to the double-well potential (with $\kappa=1$) in which the energy is above the energy barrier. In this problem 
there are two turning points on the real axis, connected by the cycle $\gamma_{123}$, and two complex turning points along the imaginary axis, 
connected by the cycle $\gamma_2$. 
The angles appearing in the TBA system are 
\be
\label{angles-quartic}
\phi_1= -\phi_{12} =-\alpha, \qquad \phi_2= \pi, \qquad \phi_{123}=0,  
\ee
where $\alpha$ depends on the energy $E$. The EQC for this potential requires two ingredients. 
First of all, it involves the WKB period associated to the cycle 
$\gamma_{123}$, which corresponds to the classical motion between the two turning points on the real axis. This is the cycle 
that one would consider in an all-orders, perturbative WKB calculation. It is related to the TBA function by, 
\be
{1\over  \hbar} s_{\pm}\left(\Pi_{\gamma_{123}}\right)(\hbar)= -\ri \epsilon_{123}\left(\theta+{\pi \ri \over 2} \pm \ri \delta\right), 
\ee
under the usual identification $\theta =-\log(\hbar)$. Note that this WKB cycle is not Borel summable, and as we discussed before, the two 
lateral resummations are related to the two infinitesimal deformations of the integration contour in the TBA system. 
Its median resummation is given, in terms of TBA functions, by 
\be
\ba
{1\over  \hbar} s_{\rm med} \left(\Pi_{\gamma_{123}}\right) (\hbar)&=m_{123}\re^{\theta} + 2\ri \int_\IR \left( {\widetilde L_1(\theta') \over \cosh(\theta-\theta' - \ri \alpha)} -
 {\widetilde L_{12}(\theta') \over \cosh(\theta-\theta' + \ri \alpha)} \right) {\rd \theta' \over 2 \pi} \\
 &-2 \, {\rm P} \int_\IR {\widetilde L_2(\theta') \over \sinh(\theta-\theta' )}{\rd \theta' \over 2 \pi}. 
 \ea
 \ee
Although in this problem there are no real instantons, there are complex instantons associated to the cycle $\gamma_2$ 
between the two complex turning points, which 
make a crucial contribution to the EQC, as noted in \cite{bpv} in a related context. The EQC reads  
\be
\label{voros-eqc}
{1\over  \hbar}  s_{\rm med} \left(\Pi_{\gamma_{123}}\right)(\hbar) -2 (-1)^n  \tan^{-1}  \left(  \re^{-{\ri  \over 2 \hbar} s\left( \Pi_{\gamma_2}\right)(\hbar)} \right) 
= 2 \pi \left( n+{1\over 2}\right), \qquad n \in \IZ_{\ge 0}. 
\ee
We have implemented this equation together with the TBA system in order to calculate the Voros spectrum of $x_n$. The 
results for $E=1$ are shown in Table \ref{table-above}. We have 
employed a discretization of the TBA equations with $2^{19}$ points. This gives an error of at most $10^{-6}$ (for the ground state) which decreases rapidly to at most $10^{-12}$ for the states 
with $n\ge 3$. 

\begin{table}[tb]
\begin{center}
  \begin{tabular}{cccc}\hline
	Level & $x_n$                    &  $E_n (x_n^{-1})$             \\ \hline
	 $0$   & $0.\, 564 \, 245\, 815$  & $1.\, 000 \, 001\, 185$                \\
	 $1$   & $1.\,424\,136\,265\,$  &$1.\, 000 \, 000 \, 021$     \\
	 $2$   & $2.\, 368\, 078\, 806$ & $1.\, 000 \, 000 \, 000$ \\  
	 $3$   & $3.\, 307 \, 532 \, 741$ & $1.\, 000 \, 000 \, 000$ \\ \hline
  \end{tabular}
  \caption{The spectrum of values of $x_n$, as computed from the exact quantization condition (\ref{voros-eqc}), and the values of $E_n (\hbar_n)$ for the Hamiltonian (\ref{h-dw}) with $E=1$.}
  \vspace{3mm}\label{table-above}
\end{center}
\end{table}

Exactly the same TBA system and the same quantization condition can be used in the case of the quartic oscillator with a 
non-negative harmonic term, e.g. with a potential of the form 
\be
V(x)= -{\kappa x^2 \over 4} + {x^4 \over 2},  \qquad \kappa\le 0. 
\ee
The case with $\kappa=0$ is the pure quartic oscillator studied in the pioneering works 
\cite{bpv,voros,voros-quartic} and revisited in the context of the ODE/IM correspondence 
\cite{dt,ddt}. As we mentioned in section \ref{sec-wc}, in this case the TBA system (\ref{6tba-red}) acquires an additional symmetry and it 
reduces to the TBA system of \cite{dt} (this has been 
already pointed out in \cite{toledo}). Indeed, in the pure quartic case one has 
\be
|m_1| =|m_{12}|={4 {\sqrt{2}} \over 3} K(\ri), \qquad |m_2|= |m_{123}|={8 \over 3} K(\ri), 
\ee
where $K$ is the elliptic integral of the first kind. The angles appearing in the kernels of the 
TBA system (\ref{6tba-red}) are as in (\ref{angles-quartic}), with $\alpha=\pi/4$. The solutions of the TBA system satisfy 
\be
\widetilde \epsilon_1= \widetilde \epsilon_{12}, \qquad \widetilde \epsilon_2= \widetilde \epsilon_{123}
\ee
with periodicity $\widetilde{\epsilon}_a(\theta)=\widetilde{\epsilon}_a(\theta+\frac{3\pi i}{2})$. The four equations (\ref{6tba-red}) reduce to two different equations for, say, $\widetilde \epsilon_a$, $a=1,2$, which can be 
easily obtained from (\ref{6tba-red}), 
\be
\ba
\widetilde \epsilon_1(\theta)&= |m_1| \re^\theta+{1\over  \pi}\int_\IR {\widetilde L_1(\theta')  \over \cosh(\theta-\theta')}  \rd \theta'+{{\sqrt{2}}\over  \pi}  \int_\IR {\cosh(\theta-\theta') \over \cosh(2 (\theta-\theta'))} \widetilde L_2(\theta') \rd \theta' , \\
\widetilde \epsilon_2 (\theta)&= |m_2| \re^\theta+{2 {\sqrt{2}} \over \pi} \int_\IR { \cosh(\theta-\theta') \over \cosh(2 (\theta-\theta'))} \widetilde L_1(\theta') \rd \theta' +{1\over  \pi}\int_\IR { \widetilde L_2 (\theta') \over \cosh(\theta-\theta')}   \rd \theta'.
\ea
\ee
This is precisely the TBA system proposed in \cite{dt} for the pure quartic oscillator. This example, together with the cubic 
example discussed in the previous section, make it clear that 
the TBA system introduced and studied in this paper provides a generalization of the ODE/IM correspondence for arbitrary polynomial potentials.

\sectiono{Conclusions and outlook}

\label{sec-conclusions}

In this paper we have provided a solution to Voros' analytic bootstrap \cite{voros-quartic} 
for arbitrary polynomial potentials in one-dimensional Quantum Mechanics. The solution takes the form 
of a TBA system for the (resummed) quantum periods which generalizes the ODE/IM correspondence \cite{dt,ddt}. 
Our approach builds upon recent progress in the solution of similar Riemann--Hilbert problems 
in e.g. \cite{gmn,gmn2,y-system,oper}. We have also illustrated our general considerations with concrete examples in 
Quantum Mechanics. In particular, we have seen how our results can be combined with 
exact quantization conditions to solve spectral problems in an efficient way. 

There are clearly many questions open by our research. Let us list some of them. 

From the TBA equations, one could identify the integrable model corresponding to the Schr\"odinger equation 
with general polynomial potential. It is important to see how the integral model ``roams'' under the wall crossing of TBA equations. It is also 
interesting to study the Schr\"odinger equation with the potential 
$q^6-\alpha q^2$ to recover the hidden $U_q(\widehat{gl}(2|1)$ structure \cite{suzuki,ddt1}. 
It was shown in \cite{blz-spectral} that the original ODE/IM correspondence of \cite{dt} can be extended to incorporate terms in the potential of the form $1/q^2$. It would be interesting to see whether our formalism can also accommodate such terms. It was also pointed out in \cite{blz-spectral,dt-beyond} that the WKB expansion of monic potentials corresponds, under the ODE/IM correspondence, to an expansion of the 
TBA system in terms of local charges of the corresponding CFT. It would be very interesting to see whether this expansion can be generalized to general polynomial potentials, in which the perturbative series involves complicated functions on moduli space. 

In section \ref{cc-pnp} we found a link between the effective central charge of the TBA system and the first quantum correction to the quantum period $\CJ$. This suggests that this period is the 
appropriate object to formulate and extend the PNP relation. In particular, it would be interesting to use these insights to find a higher genus version of the PNP relation. We also found that the effective central charge computed in section \ref{cc-pnp} is equivalent to the effective central charge of the 2d chiral algebra ${\cal A}_{-(2\ell+1)}$ in \cite{CSVY}. It would be interesting to study 
what this equivalence means in the context of the 4d/2d correspondence \cite{BLLPRR}.

Perhaps the most interesting avenue open by our results is the possibility to apply the same logic to more general problems in quantum theory. The data required by the Riemann--Hilbert problem (namely, classical limits and discontinuity structures) are provided by the resurgent structure of many quantum theories. It is then possible that one can use the logic employed here to obtain TBA equations governing the 
coupling constant dependence for other, potentially more complicated theories, including quantum field theories and string theories. 

A good starting point for these more ambitious generalizations would be 
the deformed quantum mechanics studied in \cite{gm-deformed}. The WKB periods in this theory are the quantum periods of Seiberg--Witten theory 
\cite{mirmor}, and the solution of the corresponding Riemann--Hilbert problem would provide a new perspective on the NS limit of $\CN=2$ supersymmetric 
gauge theories. One could also make a step further and consider the quantum periods of local Calabi--Yau threefolds \cite{acdkv}. Finally, 
one could try to apply the same techniques to field theory models where the resurgent structure is relatively well understood, like (complex) Chern--Simons theory 
\cite{gmp,ghatsuda}, or to field-theoretical models where semiclassical methods are reliable (see e.g. \cite{du-review} and references therein.)
 
\acknowledgments

We would like to thank Jorgen Andersen, Carl Bender, Sergio Cecotti, Santiago Codesido, Bertrand Eynard, 
Davide Gaiotto, Jie Gu, Lotte Hollands, Maxim Kontsevich, Greg Moore, Andy Neitzke, Ricardo Schiappa, Amit Sever, Junji Suzuki and Roberto Tateo for discussions 
and correspondence. We are particularly grateful to Jonathan Toledo for making available to us his unpublished note 
on wall-crossing. M.M. and H.S. would like 
to thank the conference {\it String-Math 2018} for hospitality. M.M. would like to thank the BIRS for 
hospitality during the workshop {\it Geometry and physics 
of quantum curves}, and the Erwin Schr\"odinger Institute for hospitality during the workshop {\it Geometric correspondences of gauge theories}. 
H.S. would like to thank the University of Geneva for hospitality. 
The work of K.I. is supported in part by Grant-in-Aid for Scientific Research 15K05043, 18K03643 and 16F16735 from 
Japan Society for the Promotion of Science (JSPS). The work of M.M. was supported in part by the Fonds 
National Suisse, subsidies 200020-175539 and 200020-141329, 
and by the Swiss-NSF grant NCCR 51NF40-141869 ``The Mathematics of Physics'' (SwissMAP). The work of H.S. is supported in part by JSPS Research Fellowship 17J07135 for Young Scientists, from Japan Society for the Promotion of Science (JSPS). 


\appendix

\sectiono{Perturbative calculation of the quantum periods}
\label{app-ha}

In the exact WKB method, we are often interested in computing the perturbative expansion of the WKB periods (\ref{pi-wkb-expansion}) to high order. 
Different methods have been developed in the literature in order to achieve this. In this Appendix we will present two different methods. The first 
one uses the holomorphic anomaly equations of (refined) topological string theory \cite{bcov, hk,kw}, and is based on the results of \cite{cm-ha}. The second one is based on the construction 
of differential operators.

\subsection{Holomorphic anomaly equation} 

Let us assume that the WKB curve is elliptic, so that there are only two independent cycles in the 
problem which we will call the $A$-cycle and the $B$-cycle (curves of higher genus can be also 
analyzed with this method \cite{fkn}). They define respectively the quantum periods 
$\nu$ and $\nu_D$, as in (\ref{defnus}). As shown in \cite{cm-ha}, the holomorphic anomaly 
equations of topological string theory give the expansion of the quantum free energy
\be
F^{\rm NS}(\nu) = \sum_{n \ge 0} \hbar^{2n} F_n^{\rm NS}(\nu)
\ee
as a function of $\nu$. However, in this formalism the quantum period $\nu$ is parametrized by a 
value of the modulus, which we will denote by $\xi_0$ which is related to $\nu$ by the equation defining the {\it classical} period. We then have 
\be
\nu= \Pi_A^{(0)}(\xi_0). 
\ee
At the same time, the actual value of the modulus, $\xi$, has a non-trivial expansion in $\hbar$ when expressed in terms of $\xi_0$, and one finds:
\be
\label{xi-ser}
\xi(\nu, \hbar)= \xi_0(\nu)+ \sum_{n\ge 1} \xi_n(\nu) \hbar^{2n}. 
\ee
The PNP relations make it possible to obtain this power series from the power series of $F^{\rm NS}$. Then, we have the identity:
\be
\label{identity}
\Pi_A (\xi(\nu, \hbar))= \Pi_A^{(0)}(\xi_0). 
\ee
By expanding in $\hbar$, we find equations for the $\Pi_A^{(n)}$. For example, at leading order in $\hbar^2$, (\ref{identity}) gives
\be
{\partial \Pi_A^{(0)} \over \partial \xi} \xi_1+ \Pi_A^{(1)}=0, 
\ee
and we conclude that 
\be
 \Pi_A^{(1)}=-\xi_1 {\partial \Pi_A^{(0)} \over \partial \xi}. 
 \ee
 Similarly, for the B-periods we have by definition
\be
\ri \Pi_B(\xi(\nu, \hbar)) = {\partial F^{\rm NS} \over \partial \nu}. 
\ee
From this identity we deduce for example 
\be
\ri \Pi^{(1)}_B = -\ri \xi_1 {\partial \Pi_B^{(0)} \over \partial \xi}+ {\partial F_1^{\rm NS} \over \partial \nu}. 
\ee

 The results of \cite{cm-ha} give the values of $\xi_n$, $F_n^{\rm NS}$ in terms of 
 (quasi)modular forms of the $\tau$ parameter of the WKB curve. The modular forms 
 are the Eisenstein series $E_2(\tau)$ and
 \be
 \ba
 K_2\left(q\right) &:= \vartheta_3^4\left(\tau\right) + \vartheta_4^4\left(\tau\right),\\
 K_4\left(q\right)&:=\vartheta_2^8\left(\tau\right), 
 \ea
 \ee
 Using the above results, we can compute the quantum periods $\Pi_{A,B}^{(n)}$ systematically in terms of 
 modular forms. In order to do this, it is useful to express all the ingredients of the 
 derivation in terms of modular forms as well, which is easily done by using the results of \cite{cm-ha}. 
 
 Proceeding in this way, we obtain for example, for the cubic oscillator (\ref{cubic-pot}) with $\kappa=1$:
 \be
\ba
\Pi_A^{(0)}&={\pi \over \sqrt{2}}  {2 E_2 (K_2^2 + 3 K_4) + K_2(9 K_4 -K_2^2) \over 45  (K_2^2 + 3 K_4)^{5/4}}, \\
\Pi_A^{(1)}&=\frac{\pi  \left(K_2^2+3 K_4\right)^{5/4} \left(\left(K_2^2+3 K_4\right)^2-2 E_2
   \left(K_2^3-9 K_2 K_4\right)\right)}{18 \sqrt{2} K_4
   \left(K_2^2-K_4\right)^2}, 
   \ea
\ee
For the double-well (\ref{dw-potential}) with $\kappa=1$, we find
\be
\ba
\Pi_A^{(0)}&={\sqrt{2}} \pi  {2 E_2 K_2 + 3 K_4 - K_2^2 \over 72 K_2^{3/2}},\\ 
\Pi_A^{(1)}&=\sqrt{2}  \pi \frac{  K_2^{3/2} \left(-2 E_2 K_2^2+3 E_2
   K_4+K_2^3\right)}{9 K_4 \left(K_2^2-K_4\right)}. 
   \ea
   \ee
   The $B$-periods can be easily obtained from the $A$-periods by the following rule, 
\be
 \Pi_B^{(n)}= \mu \, \tau    \Pi_A^{(n)} \bigl|_{E_2 \rightarrow E_2^D}. 
\ee
   where $\mu=1,2$ for the cubic potential and the double-well, respectively, and \cite{coms}
   \be
 E^D_2 =E_2 +{6 \over \pi \ri \tau}. 
   \ee
It is easy to see that general values of $\kappa$ can be incorporated by including a factor $\kappa^{-5n +5/2}$ in the case of the cubic well, and a factor $\kappa^{-3n+3/2}$ in the case of the double-well, and changing 
the value of $\tau$ accordingly.

\subsection{Differential operators} 
We present another approach to the calculation of  higher order corrections to the quantum periods for 
the cubic and quartic potentials \cite{huang,ito-okubo}. In this approach it is convenient to set $V(q)-E$ in (\ref{schrodinger}) as 
\be
V(q)-E=q^{r+1} -u_1 q^{r-1}-\dots - u_{r},
\ee
which is a $(r+1)$-th order polynomial in $q$. We consider the cycle on the WKB curve $\tilde{y}^2=V(q)-E$, which is slightly different from the convention used in the main text\footnote{The quantum periods will differ by some overall numerical factors.}. $\Pi^{(0)}_{\gamma}$ is  the period integral of the curve $\tilde{y}^2=V(q)-E$
\be
\Pi^{(0)}_{\gamma}=\ri \oint_{\gamma} \tilde{y} \rd q.
\ee
Since 
\be
\partial_{u_i}\tilde{y} \, \rd q =-{q^{r-i} \over 2\tilde{y}}\rd q, \qquad i=1,\cdots, r,
\ee
provide a basis of meromorphic differentials on the WKB curve,
one can expand the $p_{n}(q)\rd q$ in (\ref{pi-wkb-expansion}) in terms of this basis:
\be
p_{n}(q) \rd q=\ri \sum_{j=1}^{r} b_j^{(n)} \partial_{u_j} \tilde{y} \rd q +\partial_q(*),
\label{eq:p2n}
\ee
where the second term denotes a total derivative. It follows that the quantum corrections to the periods are given by
\be
\Pi^{(n)}_{\gamma}=\sum_{i=1}^{r} b_i^{(n)} \partial_{u_i}  \Pi^{(0)}_{\gamma} .
\ee
We can find the coefficients $b_i^{(n)} $ by solving (\ref{eq:p2n}).
Here we present the results for the cubic ($r=2$) and the quartic ($r=3$) cases up to some higher order.

Let us first consider the cubic curve given by
\be
\tilde{y}^2=q^3- u_1 q-u_2
\ee
whose discriminant is given by $\Delta=4 u_1^3-27 u_2^2$. By solving (\ref{eq:p2n}) we find that the first four corrections are given by
\be
\Pi^{(n)}=b^{(n)}_1\partial_{u_1}\Pi^{(0)}+b^{(n)}_2 \partial_{u_2} \Pi^{(0)}
\ee
with
\be
\ba
b^{(1)}_1&={9u_2\over 4\Delta}, \quad b^{(1)}_2={u_1^2\over 2 \Delta},\\
b^{(2)}_1&={21 u_1^2(308 u_1^3+5697 u_2^2)\over 160 \Delta^3}\quad 
b^{(2)}_2={63 u_1 u_2 (788 u_1^3 +2467 u_2^2)\over 320 \Delta^3},\\
b^{(3)}_1&={81 u_1 u_2(24549712 u_1^6 + 344562120 u_1^3 u_2^2 + 304562349 u_2^4)\over 17290 \Delta^5}, \\
b^{(3)}_2&={1\over 17290\Delta^5}
(69623840 u_1^9 + 5529102768 u_1^6 u_2^2 + 
   26089138530 u_1^3 u_2^4 + 6057679446 u_2^6),\\
%
   b^{(4)}_1&={1\over 573440 \Delta^7}81 u_1^2 u_2 (277770953152 u_1^9 + 11292809333904 u_1^6 u_2^2 + 
   63249377035572 u_1^3 u_2^4\\
   &
    + 47953365866955 u_2^6),
\\
b^{(4)}_2&={1\over 573440 \Delta^7}
6 (421453455808 u_1^{12} + 62854121561616 u_1^9 u_2^2 + 
   782827801494228 u_1^6 u_2^4\\
   &
    + 1375370060382675 u_1^3 u_2^6 + 
   139863680471790 u_2^8).
   \ea
\ee
In particular, at $u_1=0$, up to the 18-th order, they are found to be
\be
\ba
b_1^{(0)}&=0,\quad b^{(0)}_2={6 u_2 \over 5},\\
b^{(1)}_1&=-{1\over 12 u_2}, \quad b_2^{(1)}=0\\
b_1^{(3)}&=0,\quad  b^{(3)}_2=-{21983 \over 933120 u_2^4}\\
b^{(4)}_1&=-{26317819\over 188116992 u_2^6}, \quad b_2^{(4)}=0\\
b_1^{(6)}&=0,\quad b^{(6)}_2=-{14175476893921\over 2298682146816 u_2^9}\\
b^{(7)}_1&=-{61331166685678399 \over 391193907167232 u_2^{11}}, \quad b_2^{(7)}=0\\
b_1^{(9)}&=0,\quad b^{(9)}_2=-{2807892430989520754556859 \over 47734804593462214656 u_2^{14}}
\ea
\ee
and other $b_i^{(n)}$ for $n=2,5,8$ are zero. 
These coefficients are confirmed by the CFT approach to the T-function \cite{blz1}.

Next we consider the quartic case with the curve
\be
\tilde{y}^2=q^4 -u_1 q^2-u_2 q- u_3
\ee
whose discriminant is given by
\be
\Delta=4 u_1^3 u_2^2-27 u_2^4-16 u_1^4 u_3+144 u_1 u_2^2 u_3-128 u_1^2 u_3^2-256 u_3^3.
\ee
Solving (\ref{eq:p2n}), we find that the coefficients of the first two corrections are given by
\be
\ba
b_1^{(1)}&={4 (u_1^4-9 u_1 u_2^2+16 u_1^2 u_3+48 u_3^2)\over 3\Delta}\\
b_2^{(1)}&=0\\
b_3^{(1)}&={3 u_1^2 u_2^2-16 u_1^3 u_3 +36 u_2^2 u_3-64 u_1 u_3^2 \over 3 \Delta}
\ea
\ee
\be
\ba
b_1^{(2)}&={1\over 45\Delta^3}2 \Bigl(1792 u_1^{13} - 39768 u_1^{10} u_2^2 + 316302 u_1^7 u_2^4 - 
   1022301 u_1^4 u_2^6 + 1137240 u_1 u_2^8 + 40800 u_1^{11} u_3\\ 
   &- 
   598896 u_1^8 u_2^2 u_3 + 2354760 u_1^5 u_2^4 u_3 - 
   2599128 u_1^2 u_2^6 u_3 + 390528 u_1^9 u_3^2 - 
   3069696 u_1^6 u_2^2 u_3^2 \\
&   + 10014624 u_1^3 u_2^4 u_3^2 - 
   15688080 u_2^6 u_3^2 + 1555456 u_1^7 u_3^3 - 
   12363264 u_1^4 u_2^2 u_3^3 + 29310336 u_1 u_2^4 u_3^3 \\
   &+ 
   405504 u_1^5 u_3^4 - 9934848 u_1^2 u_2^2 u_3^4 - 
   12656640 u_1^3 u_3^5 + 74207232 u_2^2 u_3^5 - 
   23101440 u_1 u_3^6\Bigr),\\
   b_2^{(2)}&=0,
      \ea
\ee
\be
\ba
   b_3^{(2)}&={1\over 45 \Delta^3}\Bigl(-1344 u_1^{11} u_2^2 + 26676 u_1^8 u_2^4 - 185139 u_1^5 u_2^6 + 
  355752 u_1^2 u_2^8 + 1792 u_1^{12} u_3 - 
  61296 u_1^9 u_2^2 u_3 \\
   & + 611736 u_1^6 u_2^4 u_3 - 
  500904 u_1^3 u_2^6 u_3 - 2449440 u_2^8 u_3 + 
  38784 u_1^{10} u_3^2 - 958464 u_1^7 u_2^2 u_3^2 - 
  48096 u_1^4 u_2^4 u_3^2 \\
  &
  + 6792336 u_1 u_2^6 u_3^2 + 
  708096 u_1^8 u_3^3 - 1660416 u_1^5 u_2^2 u_3^3 - 
  432000 u_1^2 u_2^4 u_3^3 + 5582848 u_1^6 u_3^4\\
  & - 
  25288704 u_1^3 u_2^2 u_3^4 + 58102272 u_2^4 u_3^4 + 
  16957440 u_1^4 u_3^5 - 121614336 u_1 u_2^2 u_3^5 + 
  10321920 u_1^2 u_3^6\\
  &
   - 21626880 u_3^7\Bigr).\nonumber
\ea
\ee

In this Appendix we have presented two different approaches to compute the corrections of WKB periods.
The quantum corrections computed in these two approaches match very well, up to the overall numerical factors due to the different normalizations.

\bibliographystyle{JHEP}

\linespread{0.6}
\bibliography{biblio-ims}

\end{document}